%% file: crowd-main-arxiv.tex
\documentclass[11pt,letterpaper]{article}

\usepackage{amsmath}
\usepackage{amssymb}

\usepackage{graphicx}

\usepackage{cite}

\usepackage{color}

\usepackage[lined,algonl,boxed]{algorithm2e}
\usepackage{epsfig}
\usepackage{epstopdf}
\usepackage{graphicx}
\usepackage{subfigure}
\usepackage{multirow}
\usepackage[flushleft]{threeparttable}
\usepackage{array}
\usepackage{textcomp}
\usepackage{multirow}
\usepackage{url}

\usepackage[labelfont=bf,labelsep=period,justification=raggedright]{caption}

\date{}

\newcommand{\hide}[1]{} 
\newcommand{\vpara}[1]{\vspace{0.1in}\noindent\textbf{#1 }}

\begin{document}

\begin{flushleft}
{\Large
\textbf{Inferring Unusual Crowd Events From Mobile Phone Call Detail Records}
}
\\
Yuxiao Dong$^{\dag}$, 
Fabio Pinelli$^{\ddag}$, Yiannis Gkoufas$^{\ddag}$, Zubair Nabi$^{\ddag}$, Francesco Calabrese$^{\ddag}$, 
Nitesh V. Chawla$^{\dag}$
\\
{\bf \dag} Interdisciplinary Center for Network Science and Applications, Department of Computer Science and Engineering, University of Notre Dame, Notre Dame, IN, United States of America
\\
{\bf \ddag} IBM Research, Ireland
\\
ydong1@nd.edu, \{fabiopin, yiannisg, zubairn, fcalabre\}@ie.ibm.com, nchawla@nd.edu
\end{flushleft}

\input{abstract.tex}

\input{intro.tex}

\input{problem.tex}

\input{model.tex}

\input{exp.tex}

\input{tool.tex}

\input{related.tex}

\input{conclusion.tex}

\vpara{Acknowledgments.}
We wish to thank the Orange D4D Challenge (http://www.d4d.orange.com) organizers for releasing the data we used for testing our algorithms. Research was sponsored in part by the Army Research Laboratory under Cooperative Agreement Number W911NF-09-2-0053 and the U.S. Air Force Office of Scientific Research (AFOSR) and the Defense Advanced Research Projects Agency (DARPA) grant $\#$FA9550-12-1-0405.

\small
\bibliographystyle{abbrv}
\bibliography{references-dong}  

\end{document}

%% file: abstract.tex

\begin{quote}
\section*{Abstract}
The pervasiveness and availability of mobile phone data offer the opportunity of discovering usable knowledge about crowd behaviors in urban environments.
Cities can leverage such knowledge in order to provide better services (e.g., public transport planning, optimized resource allocation) and safer cities. Call Detail Record (CDR) data represents a practical data source to detect and monitor unusual events considering the high level of mobile phone penetration, compared with GPS equipped and open devices.
In this paper, we provide a methodology that is able to detect unusual events from CDR data that typically has low accuracy in terms of space and time resolution.
Moreover, we introduce a concept of unusual event that involves a large amount of people who expose an unusual mobility behavior.
Our careful consideration of the issues that come from coarse-grained CDR data ultimately leads to a completely general framework that can detect unusual crowd events from CDR data effectively and efficiently. 
Through extensive experiments on real-world CDR data for a large city in Africa, we demonstrate that our method can detect unusual events with 16\% higher recall and over 10 times higher precision, compared to state-of-the-art methods. 
We implement a visual analytics prototype system to help end users analyze detected unusual crowd events to best suit different application scenarios. 
To the best of our knowledge, this is the first work on the detection of unusual events from CDR data with considerations of its temporal and spatial sparseness and distinction between user unusual activities and daily routines. 

\end{quote}


%% file: intro.tex
\section{Introduction}

The ubiquity of mobile devices offers an unprecedented opportunity to analyze the trajectories of movement objects in an urban environment, which can have a significant effect on city planning, crowd management, and emergency response~\cite{Calabrese:14}.
The big data generated from mobile devices, thus, provides a new powerful social microscope, which may help us to understand human mobility and discover the hidden principles that characterize the trajectories defining human movement patterns.
Cities can leverage the results of the analytics to better provide and plan services for citizens as well as to improve their safety.
For example, during the occurrence of expected or chaotic events such as riots, parades, big sport events, concerts, the city should be able to provide a proactive response in allocating the correct amount of resources, adapt public transport services, and more generally adopt all possible actions to safely handle such events.
Many methods have been proposed in the literature to detect groups of people moving together from a trajectory database \cite{swarm,denseareas2,Convoy:VLDB08,Gathering:ICDE13}, specifically the GPS data.
However, only a very little percentage of people currently carry GPS devices, and share their movement trajectories with a central entity that can use them to identify crowd events.

In this paper, we study the problem of unusual event detection from mobile phone data that is opportunistically collected by telecommunication operators, in particular the Call Detail Records (CDR).
In 2013, the number of mobile-phone subscriptions reached 6.8 billion, corresponding to a global penetration of 96\%. The pervasiveness of mobile phones is spreading fast, with the number of subscriptions reaching 7.3 billion by 2014, from a recent report by International Telecommunications Union (ITU) at 2013 Mobile World Congress~\cite{Dong:KDD14}.
Therefore, CDR data represents a practical data source to detect and monitor unusual events considering the high level of mobile phone penetration.
This is specifically useful in developing countries where other methodologies to gather crowd movement data (e.g., GPS or cameras) are very expensive to be installed.

The task of detecting unusual events from CDR data is very different from previous work on fine-grained trajectory data, such as GPS data, and presents several unique challenges. 
\textbf{Temporal sparseness:} CDR data only records the user location when a call or text message is made or received, thus is temporally sparse since call or message frequency of users is usually low and unpredictable. 
\textbf{Spatial sparseness:} The location information of users when they make a call or message is recorded as the location of the antenna, which brings the spatial sparseness of CDR data. 
\textbf{Non-routine events:} Our objective is to detect unusual crowd events from human daily movements, which mostly consist of usual routines. Thus, it is necessary to discriminate unusual crowd movements from routine trajectories. 

To address these challenges, we aim to estimate the location of users in absence of spatio-temporal observations (i.e., the users don't make phone calls), detect groups of people moving together, and proactively discover unusual events.
We propose a general framework to infer unusual crowd events from mobile phone data.
Specifically, our contributions can be summarized as follows:
\begin{itemize}
\item We first define the cylindrical cluster to capture sparse spatio-temporal location data and provide practical methods to extract crowd events from CDR data, and further formalize the unusual crowd event detection problem by considering the similarity between individuals' trajectories and their historical mobility profiles.
\item We provide a Visual Analytics Prototype System to help the end user (e.g. a city manger or analyst) analyze the detected crowd events and set the values of the parameters to best suit an application scenario.
\item Finally, we evaluate our proposed framework on a real-world CDR dataset and demonstrate its effectiveness and efficiency. Our method significantly outperforms (10$\times$ precision and +16\% recall) previous event detection methods on GPS data with verification on real-world unusual crowd events.
\end{itemize}
The mobile phone CDR data used in this work is collected from Cote d'Ivoire over five months, from December 2011 to April 2012.
During that period, this Africa country faced the Second Ivorian Civil War and political crisis\footnote{\small\url{http://en.wikipedia.org/wiki/2010-11_Ivorian_crisis}}.
From the election of new president and parliament, continued outbreaks of post-election conflicts happened, including boycott, violence and protest etc.
The experimental results on this real-world dataset deliver the effectiveness of our proposed methodologies, which demonstrates the significant importance of our work in the supervision of unusual crowds and events for city and country management.
Moreover, through the proposed method, city mangers and officials can gain insights into non-ticketed events taking place in public spaces, which could lead to estimating the number of attendees and to estimating the event's success. A particular instance of such method has been recently implemented to help the police and the event organizers monitor visitors to the Mons 2015 - European Capital of Culture Opening Ceremony \cite{SEA:Netmob2015}.


%% file: problem.tex

\section{Unusual Event Detection Problem}
\label{sec:problem}

\begin{figure*}[!t]
\centering
  \includegraphics[width=5in]{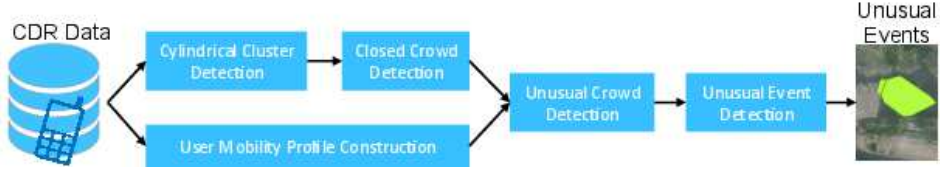}
  \caption{ Process flow of the system to detect unusual events. 
  \normalsize}
\label{fig:flow}
\vspace{-0.5cm}
\end{figure*}

Given the nature of CDR data, we face three major challenges in extracting accurate individual trajectories.
First, we can only record user locations when they make calls or receive calls (text messages).
As most mobile users do not make phone calls frequently and periodically, positions are not regularly sampled, as opposed to GPS navigation systems. Moreover, mobile users do not follow a call pattern consistently with others in the group.
Second, when a user makes a call, CDR data only records the base station she is using, providing very low quality location information.
Finally, the scenario that we are considering---e.g., going to a protest---is not consistent with an individual's daily activity pattern such as going from home to office, thus we cannot leverage the previous history of the user to enrich his trajectory to make it more accurate.

We formally define the problem of unusual event detection and decompose the problem in different steps that enable us to solve the challenges brought by CDR data. 
Figure \ref{fig:flow} shows the process flow to detect unusual events. The system receives CDR data as input, extracts clusters, and detected crowds from the sequences of clusters. Then, the system verifies some constraints for each crowd and it labels them as unusual if necessary. Subsequently, one or more unusual crowds compose unusual events.

Let $DB_{CDR} = \{call_1, call_2, \cdots, call_n\}$ denote the set of all calls collected from a mobile phone network. 
We define a call as a tuple $call_i = <t_i, v_j, l_k>$, which means a user $v_j$ makes or receives a call at location $l_k$ at timestamp $t_i$, where $v_j \in V, t_i \in T, l_k \in L$.
$V$ is the set of all users and $T$ denotes all possible timestamps.
Specifically, $l_k$ stands for the geographical location of the $k^{th}$ mobile network antenna and $L$ means the set of locations of all antennas found in $DB_{CDR}$.
We define the individual mobility trajectory~\cite{Trajectory:KDD07,Car:KDD11} for each user as follows.

\textit{Definition 1.}
\textit{Individual Trajectory:} A user $v_j$'s mobility trajectory from start time $t_p$ to end time $t_q$ is defined as a sequence of spatio-temporal tuples $s_j^{pq} = \bigcup(t_i, l_k)$, where $t_p \leq t_i \leq t_q$ and $s_j^{pq} \in S$.
$S$ stands for the set of user trajectory sequences.

\vpara{Cylindrical Cluster.} 
The first step to identify crowd events from individual trajectories is to find, at any specific timestamp, clusters of individuals that are very close in space. However, since CDR data is very sparse on the time scale (i.e., users do not make calls regularly and synchronized with each others), we propose the concept of \textit{cylindrical} \textit{cluster} in coarse-grained spatio-temporal data. 
Finer grain clustering, such as density-based clustering \cite{DBSCAN:KDD96} cannot be applied as the antenna is the lowest level of spatial resolution available in the data. 
Indeed, users are already clustered by association to the antenna they use at each call (which defines a specific coverage area in the city, ranging from a few hundred squared meters to a few kilometers). 

\textit{Definition 2.}
\textit{Cylindrical Cluster:} Given a CDR database $DB_{CDR}$ which contains individual calls with time and antenna information, and a scale threshold $\epsilon_n$, the cylindrical cluster $CC_t$ at timestamp $t$ is a non-empty subset of users $V_t \subseteq V$ satisfying the following conditions:
\begin{itemize}
\item {Connectivity.} $\forall v_i \in V_t$, $v_i$ makes at least one call by using antenna $a_x$, in the interval [$t$ - $\epsilon_t$, $t$ + $\epsilon_t$].
\item {Scale.} The number of users $|V_t|$ in $CC_t$ is no less than $\epsilon_n$.
\end{itemize}

Figure \ref{fig:prob_ex_crowd} shows an illustrative example for cylindrical clusters.
Given a timestamp $t_1$, we can see that $user 1, user 2, user 3$ and $user 4$ make calls during time interval $[t_1-\epsilon_t, t_1+\epsilon_t]$.
Also, $user 3, user 1$ and $user 2$ use the same antenna which is different from user $user 4$'s.
Then they are clustered into two groups.
One potential issue 
is that there may exist multiple locations for one single user if she/he makes multiple calls during time interval [$t_1-\epsilon_t$, $t_1+\epsilon_t$].
A number of methods can be considered to assign one single location from multiple locations, such as the central position or the most common position.
We use the most common position due to its ease of calculation and understanding.

\vpara{Crowd.}
In order to detect crowds lasting for a certain amount of time we need to consider shared characteristics between clusters detected in consecutive timestamps.

\textit{Definition 3.}
\textit{Crowd:} Given a CDR database $DB_{CDR}$ with individual trajectories, 
a lifetime threshold $\epsilon_{lt}$, a consecutive intersection threshold $\epsilon_{ci}$ and a commitment probability threshold $\epsilon_p$, a crowd $C$ is a sequence of consecutive cylindrical  clusters $\{CC_{t_m}, CC_{t_{m+1}}, \cdots, CC_{t_n} \}$ which satisfy the following constraints:
\begin{itemize}
\item {Movement.} The number of total locations in one crowd is more than one. 
\item {Durability.} The lifetime of $C$, $C.lt$, namely the number of consecutive clusters, is greater than $\epsilon_{lt}$, i.e., $C.lt \geq \epsilon_{lt}$ where $C.lt = n-m+1$.
\item {Commitment.} At least $\epsilon_{ci}$ users appear in each cylindrical  cluster with existence probability $\epsilon_{p}$.
\end{itemize}
The movement and durability characterizations specify the types of crowd we are interested in. 
The commitment instead characterizes the fact that a certain subset of users needs to participate to all clusters.
Again, due to the spatio-temporal sparsity of the CDR data, the computation of the commitment of an user requires some further considerations.
Therefore, we propose the concept of existence probability, which is designed to overcome CDR sparsity. Indeed, as an individual is not constantly making calls, consecutive timestamps could not see all users in the cluster making calls.

We design the existence probability of one user locating in a cluster at timestamp $t$ as the proportion of the number of users in $CC_{t}$ to the number of users in $CC_{t-1}$.
The intuition for the definition of existence probability is that the user has conformity to follow others in the group that she or he was assigned to~\cite{KDD13:conformity}. 
For example, the existence probability of $user3$ in Figure \ref{fig:prob_ex_crowd} at time $t_2$ is $1/3$.
In timestamp $t_1$, $user1$, $user2$, and $user3$ stay in cluster $CC_{t_1}$, and one of them, user $user 2$, goes to cluster $CC_{t_2}$ at timestamp $t_2$.
$user 1$ and $user 3$ do not make calls in timestamp $t_2$, which results in the uncertainty of their locations.
Thus, we assign them the probability to stay with $user 2$, which is in cluster $CC_{t_2}$.
Furthermore, we make the existence probability decay over time, i.e., if a user does not appear in consecutive timestamps, such as user $user 4$ in timestamp $t_3$ and $t_4$. 
Her existence probabilities in Figure \ref{fig:prob_ex_crowd} are $[0, 1, \frac{1}{2}, \frac{1}{2}\times\frac{2}{3}]$ at each timestamp, respectively. 


Considering that a crowd is a sequence of clusters, we use the standard terminology of sequential pattern mining and affirm that: a crowd $C$ is called a closed crowd if it has no super crowds, which means there does not exist super sequences containing $C$. 

\begin{figure}[!t]
\centering
\subfigure[\scriptsize Cylindrical Cluster]{
\label{fig:prob_ex_crowd}
\includegraphics[width=2.8in]{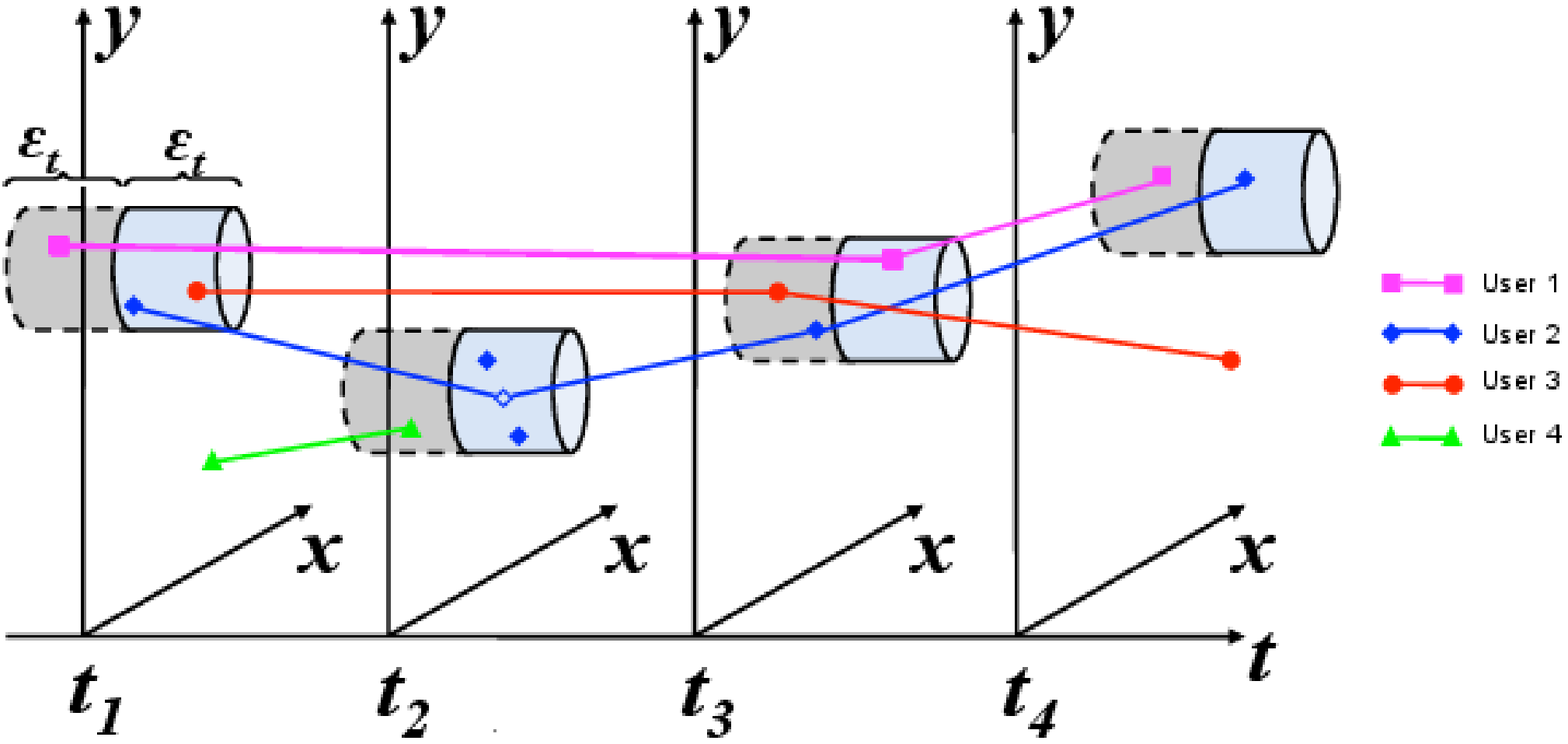}
}
\subfigure[\scriptsize Closed Crowd]{
\label{fig:ex_closeprop}
\includegraphics[width=2.8in]{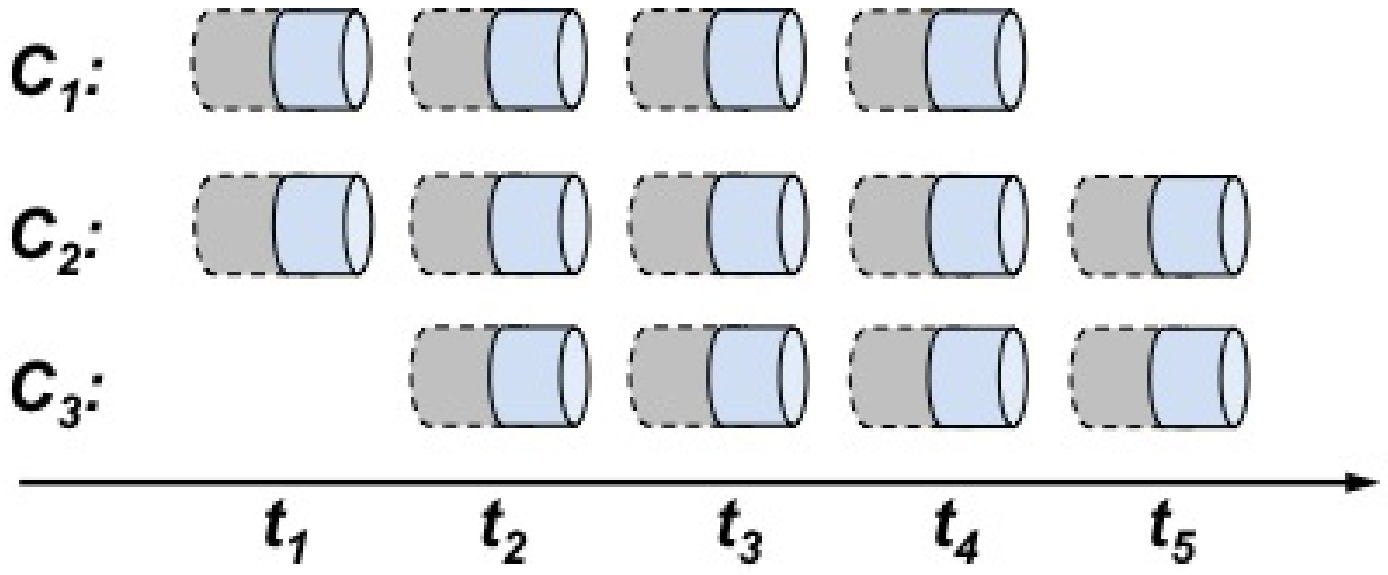}
}
\caption{\label{fig:examples} Illustrative Examples of Cylindrical Cluster and Closed Crowd.}
\end{figure}

\hide{

\begin{figure}[!t]
\centering
  {\includegraphics[width=3.2in]{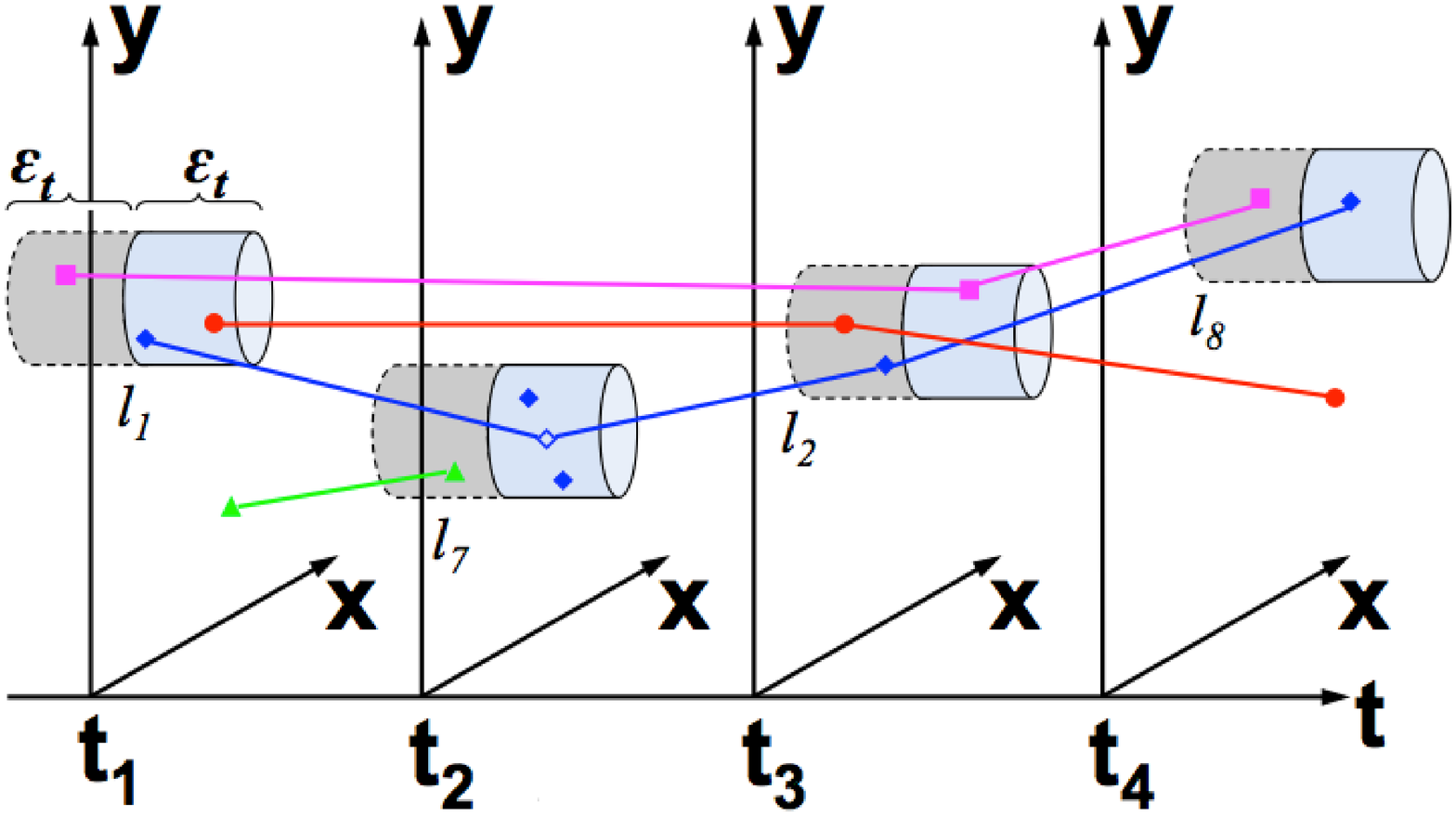}}
  \caption{ Illustrative Example of Cylindrical Cluster. \normalsize}
\label{fig:prob_ex_crowd}
\vspace{-0.5cm}
\end{figure}

\begin{table}[!t]
\caption{Existence probability for users in Figure \ref{fig:prob_ex_crowd}.}
\centering
\begin{tabular}{|l|l|l|l|l|l|l|l|l|}
\hline
             &   $t_1$       & $t_2$         & $t_3$             & $t_4$     \\ \hline
user $1$  & 1             & $\frac{1}{3}$ & 1                 & 1         \\ \hline
user $2$  & 1             & 1             & 1                 & 1         \\ \hline
user $3$   & 1             & $\frac{1}{3}$ & 1                 & 0         \\ \hline
user $4$ & 0             & 1             & $\frac{1}{2}$         & $\frac{1}{2}$ $\times$ $\frac{2}{3}$         \\ \hline
\end{tabular}
\label{tb:prob_ex_exist_probability}
\end{table}
}

\vpara{Unusual Crowd.} 
Usually, people have their own mobility trajectories in daily lives, such as going from home to work place everyday.
When people go to attend a concert or a protest, their trajectories differ from their usual ones.
The definition of crowd given above includes both usual daily trajectories (e.g., commuting) as well as unusual event trajectories (e.g., protests). 
This is, for instance, what the method in \cite{Gathering:ICDE13} aims to do. As we will show in the experiments section, such method generates an enormous amount of events, as opposed to what a city would need in order to identify specific unusual events. 
Here we define the concept of mobility profile to capture people's normal movement behaviors, by comparing with which we can detect abnormal mobility behaviors.

\textit{Definition 5.}
\textit{Mobility Profile:} Given a CDR database $DB_{CDR}$ with individual trajectories, one's mobility profile is the groups of locations she/he visited for each time unit (hour) in every day. Notice that a location here corresponds to an antenna.

\hide{
\begin{table}[t]
\caption{Mobility Profile for $user 4$ in Figure \ref{fig:prob_ex_crowd}.}
\centering
\begin{tabular}{|l|l|l|l|l|l|l|l|}
\hline
$t_1$     & $t_2$                & $t_3$                & $t_4$                     & $t_5$    & $t_6$      \\ \hline
$l_{2}$:3 & $l_{7}$:2, $l_{4}$:5 & $l_{9}$:4, $l_{2}$:1 & $l_{8}$:2, $l_{1}$:1      & $l_4$:3  & $l_9$:2    \\ \hline
\end{tabular}
\label{tb:prob_profile}
\end{table}

For example, setting $\alpha_f = $ 1 hour and $\alpha_p = $ 1 week, we record the locations one user stayed for every hour in each week. An illustrative example for $user4$ in Figure \ref{fig:prob_ex_crowd} is shown in Table \ref{tb:prob_profile}. In this table, we can see that $user4$ locates at position $l_{2}$ three times on average in timestamp $t_1$ every day.
In this work, we set $\alpha_f = $ one hour and $\alpha_p = $ one day. Now, we can define the unusual crowd concept.

}

\textit{Definition 6.}
\textit{Unusual Crowd:} Given the mobility profiles of users, a similarity threshold $\epsilon_{si}$, a closed crowd $C$ is said to be an unusual-crowd $UC$ if the average similarity between the trajectory of each user in the crowd and her/his mobility profile in corresponding time intervals is less than $\epsilon_{si}$.


\vpara{Unusual Event Detection.} 
Due to the inaccuracy of CDR data and to the introduction of the existence probability concept, it is possible that two or more crowds share users and thus they represent the same event. Moreover, it is possible that many crowds might correspond to the same large event (e.g., two parades converging to the same square). To group together these unusual crowds, we define the concept of unusual event:

\textit{Definition 7.}
\label{def:event}
\textit{Unusual Event:} Given two unusual crowds $UC_i$ and $UC_j$, 
$UC_i$ and $UC_j$ are connected into one unusual event if they satisfy the following principles:
\begin{itemize}
\item {Overlapping:} The ending time $C_i.t_{end}$ of crowd $C_i$ is temporally close to the beginning time $C_j.t_{begin}$ of other crowd $C_j$, w.r.t. $C_j.t_{begin} < C_i.t_{end}$.
\item {Sharing:} The number of common users, ${|C_i\bigcap C_j|}$, is larger than or equal to half of the total users ${|C_i\bigcup C_j|}$.
\end{itemize}
An unusual event is a set of unusual crowds $E = \{UC_1, UC_2, \cdots, UC_n \}$ in which any two unusual crowds are connected to each other by a path. Here one separate  unusual-crowd is also an unusual event, if it does not connect with others. 
Based on the discussed concepts above, we formalize the unusual event detection problem as follows.

\textbf{Problem 1.}
\textbf{\textit{Unusual Event Detection:}} Given all detected crowds during the interval of two timestamps, the goal of unusual event detection is to extract all unusual events happening in the time interval.

Unusual crowd event detection in mobile phone CDR data faces several unique challenges. 
First, the sparseness of CDR data comes from not only the fact that a user's location is recorded only when a call is made but also the way that this location is approximated as the cover area of an antenna that is being used by this call. 
To solve the temporal and spatial sparseness of CDR data, we propose to define user existence probability that can overcome the fact that a user's location is recorded only when a call is made, and also to leverage the idea of cylindrical cluster to address the coarseness of user locations as they are recorded as the cover area of involved antenna. 
Moreover, the problem is targeted at inferring unusual events rather than people daily routines. 
To achieve so, we propose the concept of mobility profile to distinguish unusual crowding behavior from daily movements.

%% file: model.tex
\section{Unusual Event Detection Framework}
\label{sec:model}

Given the formal definitions above, we describe now an innovative and efficient framework to detect unusual crowd events from CDR data.
Our framework is composed of four parts: cylindrical cluster detection, closed crowd detection, unusual crowd detection, and unusual event detection. 

\vpara{Cylindrical Cluster Detection.} 
Given the database of the individual calls with the respective time and antenna information, a duration threshold $\epsilon_t$, and a scale threshold $\epsilon_n$, the Cylindrical Cluster Detection algorithm maintains at each timestamp $t$ the set of users observed from each antenna $a$, in the time interval [$t$ - $\epsilon_t$, $t$ + $\epsilon_t$]. Then, for each timestamp it returns all the set of users whose size is larger than $\epsilon_n$. All the detected cylindrical clusters are stored in $ClusterDB$.

\hide{
\begin{figure}[!t]
\centering
  {\includegraphics[width=8cm]{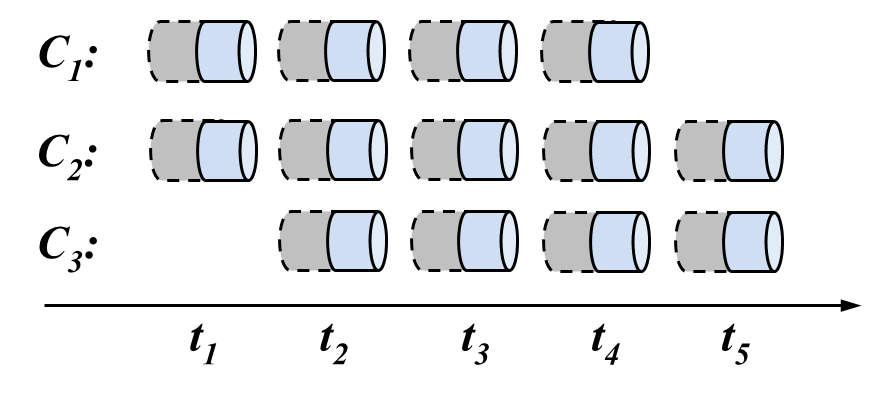}}
  \caption{ Illustrative Example of Closed Crowd. \normalsize}
\label{fig:ex_closeprop}
\end{figure}

}

\hide{
\begin{algorithm}[!t]
\caption{Closed Crowd Detection.}\label{arg:crowd}
\KwIn{\\ \ \ \ \ Cylindrical Clusters: $ClusterDB$.}
\KwOut{\\ \ \ \ \  Closed Crowds: $CCrowds$.}
\BlankLine
Initialize: CaCrs $\leftarrow \emptyset$\;
\For{t $\leftarrow$ 0 to $t_{max}$}
{
	$CCrs_{t-1} \leftarrow \emptyset$\;
	\ForEach{ Cr in CaCrs}{
		CaCls = $CSearch$(Cr, ClusterDB[t]) \;
		\eIf{CaCls == $\emptyset$}{
			\If{Cr.lt $\geq$ $\epsilon_{lt}$ \& Cr.loc $\geq$ $\epsilon_{loc}$ \& isClose(Cr, $CCrs_{t-1}$)}{
			$CCrs_{t-1}$.$add$(Cr)\;
			}
		}{
			\ForEach{Cl in CaCls}{
				Cr' $\leftarrow$ Cr + Cl\;
				CaCrs.$add$(Cr')\;
			}
		}
		$remove$ Cr\;
	}
	CCrowds $\leftarrow CCrowds \bigcup CCrs_{t-1}$ \;
	\ForEach{Cl in ClusterDB[t]}{
		\If{Cl.users.$size$() $\geq$ $\epsilon_n$}{
			Cr" $\leftarrow$ Cl\;
			CaCrs.$add$(Cr")\;
		}
	}
}
\Return{CCrowds}\;
\end{algorithm}
}

\vpara{Closed Crowd Detection.} 
The input for crowd detection is a set of cylindrical clusters $ClusterDB$ extracted at each timestamp. 
There are three constraint thresholds considered in our crowd definition: $movement$, $durability$, $commitment$.
Explicitly, if the subcrowd of one crowd meets the $durability$ and $movement$ constraints, it will satisfy the $commitment$ constraint also.
Thus the crowd definition satisfies the requirement of downward closure property, and then it is unnecessary to output all crowds, including the subcrowds of closed crowds.
To avoid the redundancy resulted from outputting subcrowds, we can follow the Lemma below 
to decide if a crowd is closed or not.
\textit{Lemma.} 
\label{lem:closecrowd}
A crowd $C$ with clusters $\{CC_{t+m}$, $CC_{t+m+1}$,$\cdots$, $CC_{t+n}\}$ is a closed crowd, if there does not exist $CC_{t+m-1}$ or $CC_{t+n+1}$ that can be added to crowd $C$ such that a new crowd is formed.
The restriction of closed crowd contains two conditions, one is that no suffixed cluster can be appended into it and the other is that no prefixed cluster can be merged in its front. 
To discover closed crowds in cluster database at current timestamp $t$, the first condition is easy to check: if there exist clusters in next timestamp $t+1$ that can be appended to current crowd $C$, then the process will continue; if not, we only need to verify whether current crowd $C$ is the subcrowd of crowds formed at current timestamp $t$.
It is not necessary to check every crowd at previous timestamps because that current crowd at timestamp $t$ can only be the subcrowd of crowds ending at timestamp $t$. 

Figure \ref{fig:ex_closeprop} shows an illustrative example for this process.
Suppose that crowds $C_1$ and $C_2$ are found as closed crowds, if there is no cluster at timestamp $t_6$  that can be appended to crowd $C_3$, then we need to further check whether it is the subcrowd of previous crowds.
It is obvious that it is impossible for $C_3$ to be the subcrowd of crowds ending at $t_4$ or earlier timestamps, such as $C_1$, but it is possible to be the subcrowd of crowds ended at $t_5$, such as $C_2$.

To find all closed crowds in $ClusterDB$, we start with iterating each timestamp in an increasing order.
At each timestamp $t$, we check whether each candidate crowd at timestamp $t-1$ can be appended by clusters at timestamp $t$. 
If the candidate crowd satisfies the \textit{movement} and \textit{durability} constraints, and at the same time it is not the subcrowd of crowds ending at timestamp $t-1$, then we can output the current candidate as a closed crowd. The current candidate crowd can then be appended by one more cluster to form a new candidate crowd at $t$. 
The candidate crowd set contains all crowds which can be appended by a new cluster at $t$.
Then we put all clusters at timestamp $t$ to it to form a new candidate crowd set at $t$.
This order of adding candidate crowd to candidate set guarantees that we only need to check whether the potential crowd is the subcrowd of closed crowds ending at the same timestamp.

\textit{Complexity}: The extraction of closed crowds is similar to the extraction of closed frequent sequential patterns whose complexity in the worst case can be approximated with $\mathcal{O}(|A|^2*|T|)$ where $|A|$ is the number of antennas (i.e. clusters) and $|T|$ is the number of timestamps.

\hide{
\begin{algorithm}[!t]
\caption{Candidate Cluster Search.}\label{arg:csearch}
\KwIn{\\ \ \ \ \ Clusters $Cls$ at $t$, candidate crowd $Cr$ at \textit{t-1}. }
\KwOut{\\ \ \ \ \  Candidate clusters: $CaCls$.}

\BlankLine
Initialize: CaCls $\leftarrow \emptyset$\;
\ForEach{Cl in Cls}{
	\If{Cl.$size$() $\geq$ $\epsilon_{n}$}{
		conusers $\leftarrow$ Cr.clusters[t-1].users $\bigcap$ Cl.users\;
		cotusers $\leftarrow$ conusers\;
		prob = conusers.$size$() / Cr.clusters[t-1].users.$size$()\;
		\If{prob $\geq \epsilon_p$}{
			\ForEach{user in Cr.users$\backslash$Cl.users}{
				$eprob_{t}$ = user.$eprob_{t-1}$ $\times$ prob\;
				\If{$eprob_{t} \geq \epsilon_p$}{
					cotusers.$add$(user)\;
				}
			}
			\If{cotusers.$size$() $\geq \epsilon_{ci}$}{
				CaCls.$add$(Cl)\;
			}
		}
	}
}
\Return{CaCls}\;
\end{algorithm}
}

\vpara{Unusual Crowd Detection.} 
With the detected closed crowds, we further verify whether their users present unusual or regular behaviors. 
As introduced in Section \ref{sec:problem}, we use mobility profile to decide whether users' movement trajectories are unusual.

To generate the mobility profiles, we scan the historic CDR data once to record the specific locations a user visited 
at each timestamp during every time period. 
For example, 
\textit{user}4's existence probability vector in corresponding crowd is $\textbf{w}_c = [0, 1, \frac{1}{2}, \frac{1}{2}\times\frac{2}{3}]$ in Figure \ref{fig:prob_ex_crowd}. 
His profile vector is extracted from his mobility profile at corresponding timestamps (from $t_1$ to $t_4$), i.e. $\textbf{w}_m = [\frac{0}{3}, \frac{2}{2+5}, \frac{1}{4+1}, \frac{2}{2+1}]$. 
There are several ways to define the similarity between user's mobility profile and his trajectory in the crowd.
We use cosine similarity to calculate the similarity score, because of its ease of understanding and implementation.
The cosine similarity between two vectors $\textbf{w}_c$ and $\textbf{w}_m$ is defined as:
$CosSim(\textbf{w}_c, \textbf{w}_m) = \frac{\textbf{w}_c \cdot \textbf{w}_m}{\| \textbf{w}_c \| \| \textbf{w}_m\|}.$ 

The Unusual Crowd Detection algorithm first calculates for each user in the crowd the similarity between her trajectory and her own mobility profile.
Then the similarities obtained are averaged, and the obtained value is greater than the $similarity$ parameter $\epsilon_{si}$, it is an unusual crowd.

\textit{Complexity}: The mobility profile construction requires a scan of the dataset, therefore its complexity is $\mathcal{O}(DB_{CDR})$. The detection of Unusual-Crowds requires for each crowd the computation of the cosine similarity for all the users being part of a crowd, thus its complexity is $\mathcal{O}(|C|*|V|)$ where $|C|$ is the number of crowds and $|V|$ the number of users.

\vpara{Unusual Event Detection.} 
With discovered unusual crowds, we finally detect their relationships and connect them into one event if they meet the requirements of Definition 7. 
In this step, we use graph theory to find and generate unusual events.
First if two unusual crowds satisfy both $overlapping$ and $sharing$ principles, we create an edge to connect them.
With this generated graph, where each node is one unusual crowd and an edge indicates that two crowds belong to the same event, the event detection is to generate all components in the graph.
Note that this graph may not only be disjoint but also include single nodes.
Each component or single node is an unusual-event that is our final goal of this work.
The first part of this algorithm checks if two unusual crowds can be connected to each other by parameters $overlapping$ and $sharing$.
The second part generates all the components in the unusual crowds graph, where any graph algorithm can be used.
The detected event contains the users in each cluster and its corresponding timestamp and location.

\textit{Complexity}: The detection of Unusual-Events requires a pair-wise comparison between all the Unusual Crowds, therefore the complexity of this procedure is $\mathcal{O}(|UC|^2)$ where $|UC|$ is the number of Unusual Crowds.

%% file: exp.tex
\section{Experiments}
\label{sec:experiments}


\subsection{Experimental Setup}
\label{sec:data}

\vpara{CDR Data.} 
The D4D Orange challenge 
made available data collected in Cote d'Ivoire over a five-month period, from December 2011 to April 2012. 
The datasets describe call activity of 50,000 users chosen randomly in every 2-week period. 
Specifically, the data contains the cell phone tower and a timestamp at which the user sent or received a text message or a call in the form of tuple \texttt{<UserID, Day, Time, Antenna>}.
Each antenna is associated with location information. 
To avoid privacy issues, the data has been anonymized by D4D data provider. 

From the CDR data, we find that about 63\% users do not make calls in consecutive hours and 19\% users make calls in only two consecutive hours. 
The pattern demonstrates the necessity of existence probability for user's location estimation, as most of users do not make regular and consecutive calls at each timestamp.
We also observe that the probability that there is one hour between one user's two calls is more than 75\% and that is 8\% for two-hour interval.
In total, there are more than 80\% two consecutive calls whose intervals are at most two hours.
These observations demonstrate the challenges of spatio-temporal sparseness on CDR data, which makes the design for degenerative existence probability reasonable for the coarse grained CDR data.

\hide{
\begin{figure}[t]
\centering
\subfigure[]{
\hspace{-0.2in}
\label{figsub:data-poshour}
\includegraphics[width=1.6in]{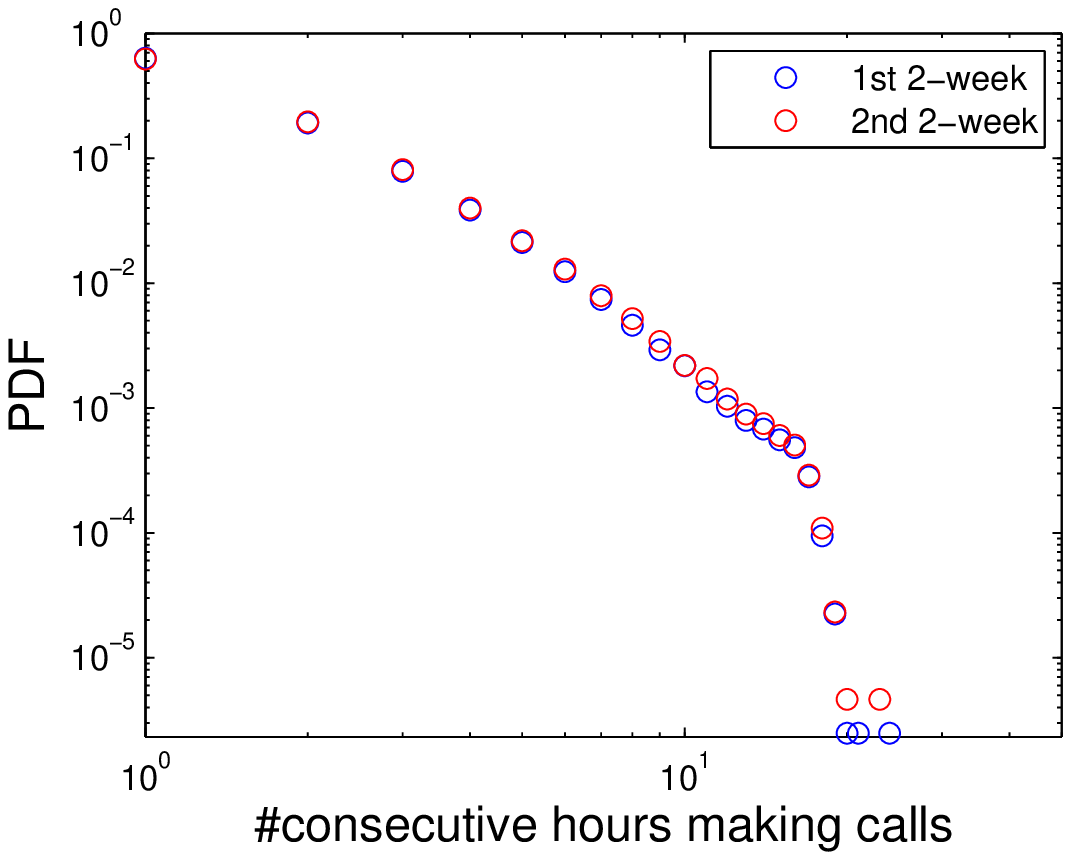}
}
\hspace{-0.2in}
\subfigure[]{
\label{figsub:data-neghour}
\includegraphics[width=1.6in]{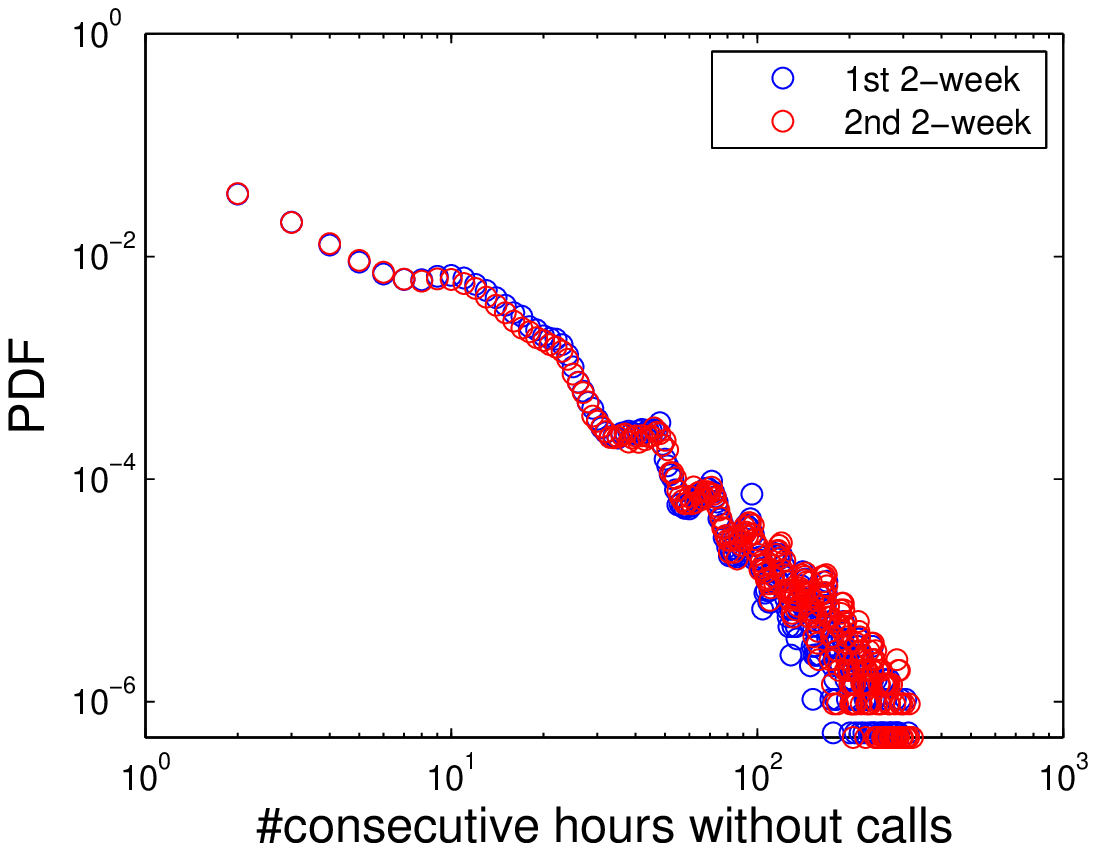}
\hspace{-0.2in}
}
\caption{\label{fig:data} (a) Distribution of consecutive hours making calls. (b) Distribution of consecutive hours without calls. }
\end{figure}
}

\vpara{Comparison Methods.} 
To the best of our knowledge, this is the first work to detect unusual crowds and events in spatio-temporal data, and it is also the first time that we discover moving clusters in CDR data. 
We compare the results of our approach with the proposed methods described in \cite{Gathering:ICDE13} (GAT) and in \cite{movingClusters} (MOV), as the methods employed in these work are also able to identify moving crowds. 
However, those methods have not been designed to work on CDR, but have to be adapted to perform the comparison. 
GAT defines a method to detect the gatherings in a trajectory dataset. A gathering is a sequence of spatial clusters with a certain number of committed users being member of an enough number of clusters. 
We use the same setting with GAT for parameters that indicate the same physical meanings in both methods. 
Clearly by following our intuition and goal of problem design, there should not exist any crowd or event at most days. 
Based on the results of parameter analysis in Section \ref{sec:efficeity-para} and the developed Visual Analytics System, we selected the following parameters $\epsilon_n$=20, $\epsilon_{lt}$=4, $\epsilon_{ci}$=10, $\epsilon_p$=0.2, and $\epsilon_{si}$=0.2.  
Since they correspond to a probability to find unusual-crowds in an hour to be around 10-15\%, which helps us focus on rare events (as opposed to business as usual events).

\begin{table*}[!t]
\caption{Comparison of the unusual event detection (UE) and gathering detection (GAT)~\cite{Gathering:ICDE13}. 
}
\centering
{\scriptsize
\begin{tabular}{|l||l||l||c||c||c||c|}
\hline
Period & Date & Event Name~\cite{netmobevents} & UE & $\#$UE & GAT & $\#$GAT\tabularnewline
\hline
\hline
\multirow{3}{*}{Dec. 05 - Dec. 18} & Dec. 07  & Anniversary of Felix Death  & $\surd$ & \multirow{3}{*}{20} & $\surd$ & \multirow{3}{*}{287}\tabularnewline
\cline{2-4} \cline{6-6}
 & Dec. 11 & Parliament election  & $\surd$ &  & $\times$ & \tabularnewline
\cline{2-4} \cline{6-6}
 & Dec. 17 & Violence & $\surd$ &  & $\surd$ & \tabularnewline
\hline
\hline
\multirow{3}{*}{Dec. 19 - Jan. 01} & Dec. 25 & Christmas day & $\times$ & \multirow{3}{*}{36} & $\times$ & \multirow{3}{*}{56}\tabularnewline
\cline{2-4} \cline{6-6}
 & Dec. 31 & New year eve & $\surd$ &  & $\surd$ & \tabularnewline
\cline{2-4} \cline{6-6}
 & Jan. 1 & New year day & $\surd$ &  & $\times$ & \tabularnewline
\hline
\hline
\multirow{2}{*}{Jan. 02 - Jan. 16} & Jan. 08  & Baptism of Lord Jesus & $\surd$ & \multirow{2}{*}{31} & $\surd$ & \multirow{2}{*}{176}\tabularnewline
\cline{2-4} \cline{6-6}
 & Jan. 14 & Arbeen Iman Hussain  & $\surd$ &  & $\times$ & \tabularnewline
\hline
\hline
\multirow{2}{*}{Jan. 17 - Jan. 29} & Jan. 17 & Visit of Hilary Clinton  & $\surd$ & \multirow{2}{*}{15} & $\surd$ & \multirow{2}{*}{481}\tabularnewline
\cline{2-4} \cline{6-6}
 & Jan. 18 & Visit of Kofi Annan  & $\surd$ &  & $\surd$ & \tabularnewline
\hline
\hline
\multirow{7}{*}{Jan. 30 - Feb. 12} & Jan. 30 & ACNF 2012 vs Angola & $\surd$ & \multirow{7}{*}{58} & $\surd$ & \multirow{7}{*}{310}\tabularnewline
\cline{2-4} \cline{6-6}
 & Feb. 04 & ACNF 2012 vs Equatorial Guinea & $\surd$ &  & $\surd$ & \tabularnewline
\cline{2-4} \cline{6-6}
 & Feb. 04 & Mawlid an Nabi Sunni  & $\surd$ &  & $\surd$ & \tabularnewline
\cline{2-4} \cline{6-6}
 & Feb. 05 & Yam & $\surd$ &  & $\times$ & \tabularnewline
\cline{2-4} \cline{6-6}
 & Feb. 08 & ACNF 2012 Semi Final VS. Mali & $\times$ &  & $\surd$ & \tabularnewline
\cline{2-4} \cline{6-6}
 & Feb. 09 & Mawlid an Nabi Shia  & $\surd$ &  & $\surd$ & \tabularnewline
\cline{2-4} \cline{6-6}
 & Feb. 12 & ACNF 2012 Final VS. Zambia  & $\surd$ &  & $\times$ & \tabularnewline
\hline
\hline
\multirow{2}{*}{Feb. 13 - Feb. 26} & Feb. 13 & Post African Cup of Nations Recovery  & $\surd$ & \multirow{2}{*}{52} & $\surd$ & \multirow{2}{*}{152}\tabularnewline
\cline{2-4} \cline{6-6}
 & Feb. 22 & Ash Wednesday  & $\surd$ &  & $\surd$ & \tabularnewline
\hline
\hline
Feb. 27 - Mar. 10 & None &  &  & 26 &  & 269\tabularnewline
\hline
\hline
\multirow{2}{*}{Mar. 11 - Mar 25} & Mar. 12 & Election of National Assembly President  & $\surd$ & \multirow{2}{*}{17} & $\surd$ & \multirow{2}{*}{342}\tabularnewline
\cline{2-4} \cline{6-6}
 & Mar. 13 & Election of National Prime Minister  & $\surd$ &  & $\surd$ & \tabularnewline
\hline
\hline
\multirow{2}{*}{Mar. 26 - Apr 08} & Apr. 01-04 & Education International Congress  & $\surd$ & \multirow{2}{*}{75} & $\surd$ & \multirow{2}{*}{1220}\tabularnewline
\cline{2-4} \cline{6-6}
 & Apr. 06 & Good Friday  & $\surd$ &  & $\surd$ & \tabularnewline
\hline
\hline
\multirow{2}{*}{Apr. 09 - Apr. 22} & Apr. 09 & Easter Monday & $\surd$ & \multirow{2}{*}{10} & $\surd$ & \multirow{2}{*}{33}\tabularnewline
\cline{2-4} \cline{6-6}
 & Apr. 13-14 & Assine fashion days  & $\surd$ &  & $\surd$ & \tabularnewline
\hline
\hline
Total  & \multicolumn{2}{c||}{} & 23/25 & 340 & 19/25 & 3326\tabularnewline
\hline
\hline
Precision & \multicolumn{2}{c||}{} & \multicolumn{2}{c||}{\textbf{0.0676}} & \multicolumn{2}{c|}{\textbf{0.0057}}\tabularnewline
\hline
\hline
Recall & \multicolumn{2}{c||}{} & \multicolumn{2}{c||}{\textbf{0.9200}} & \multicolumn{2}{c|}{\textbf{0.7600}}\tabularnewline
\hline
\end{tabular}
}
\label{tb:exp-events}
\end{table*}

\begin{figure}[!t]
\centering
\subfigure[\scriptsize Number of unusual events]{
\label{figsub:exp-weeks0-ue}
\includegraphics[width=2.8in]{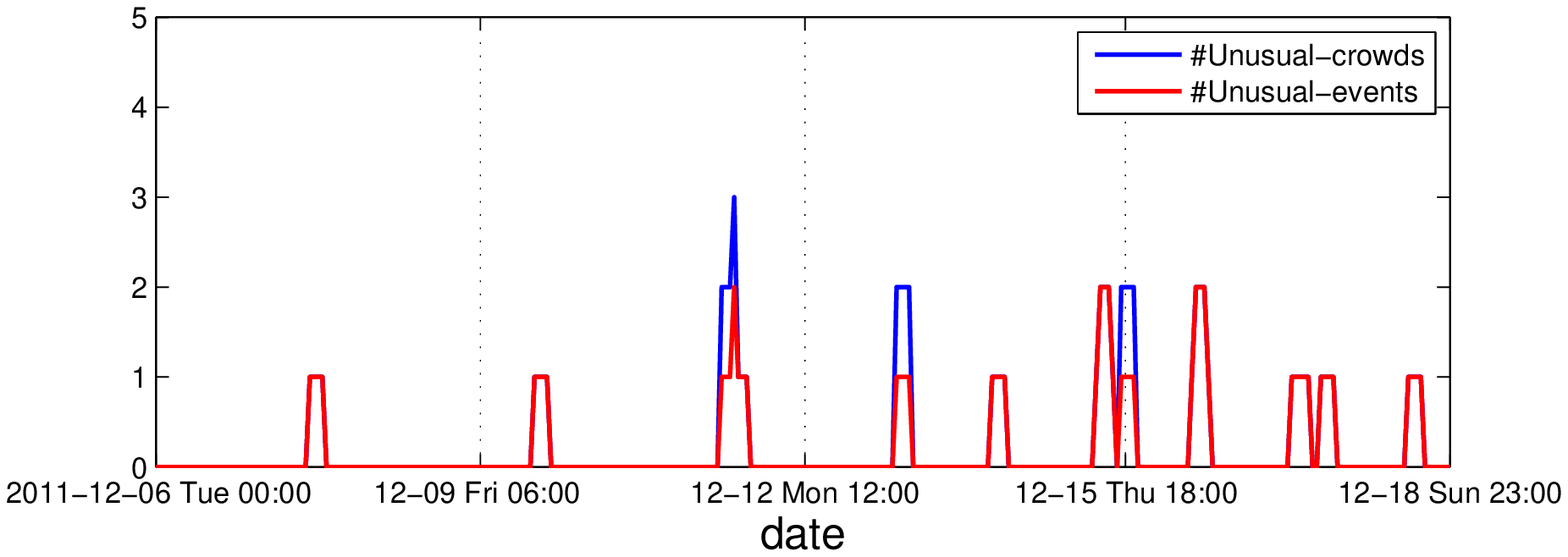}
}
\vspace{-0.1in}
\subfigure[\scriptsize Number of gatherings]{
\label{figsub:exp-weeks0-gat}
\includegraphics[width=2.8in]{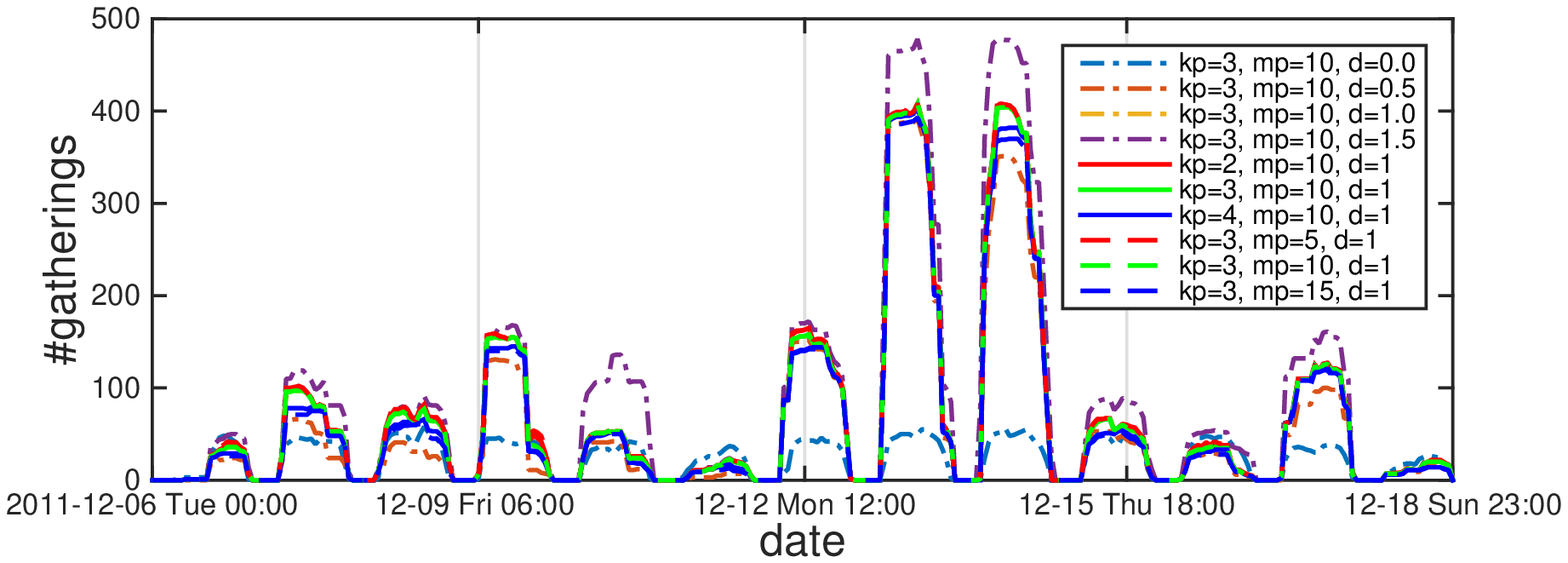}
}

\subfigure[\scriptsize Number of total call users/times ]{
\label{figsub:exp-weeks0-times}
\includegraphics[width=2.8in]{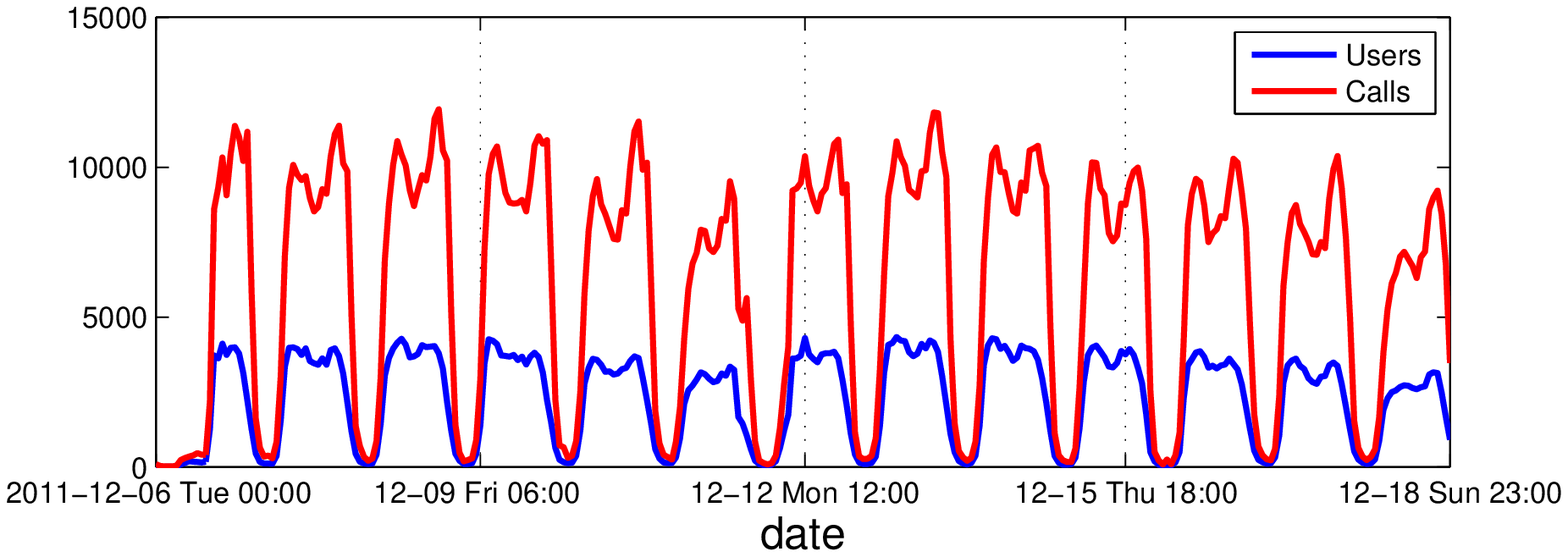}
}
\vspace{-0.2in}
\subfigure[\scriptsize Number of calls at antennas]{
\label{figsub:exp-weeks0-bids}
\includegraphics[width=2.8in]{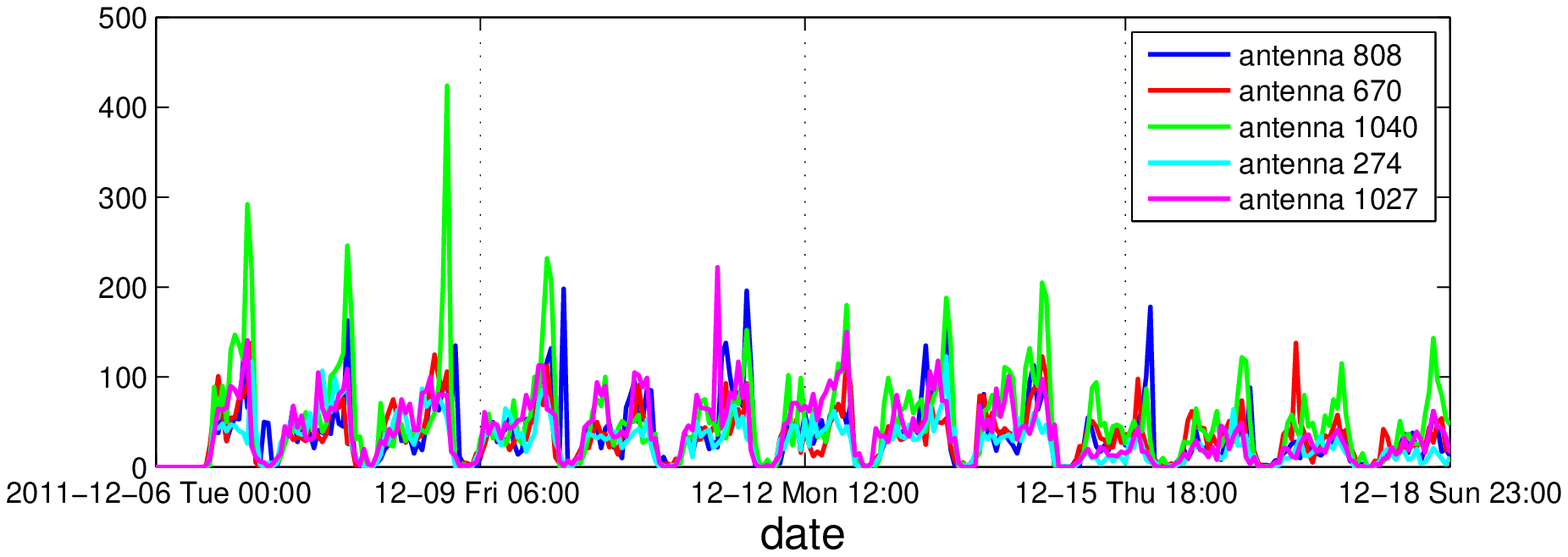}
}
\caption{\label{fig:exp-weeks0} Time series of unusual events, gatherings, users, and calls in the first two-week.}
\end{figure}

\subsection{Experimental Results}


\vpara{Detected unusual events.}
Table \ref{tb:exp-events} reports a series of events occurred in Abidjan in the different periods covered by the datasets.
In order to perform a fair study of the effectiveness of our method in comparison with GAT, we selected a third part set of events reported in \cite{netmobevents}. 
To limit the explosion in the number of detected gatherings, we have set the most restrictive values for the remaining parameters: $d=0.0$, $k_p=2$, and $m_p=5$.
Moreover, we report the total number of generated events by both methods for each two-week period and subsequently generate \emph{Precision} and \emph{Recall} scores for both algorithms.
It is possible to notice that our method detects a lower number of unusual events w.r.t. GAT. This is reflected in a higher value of Precision. Although, our method reports a lower number of events, it is able to detect a greater number of ground truth events, and this corresponds to a higher value of Recall.
Notice that the two measures represent an estimation of precision and recall since the ground truth is not given.
Indeed the list of events in \cite{netmobevents} is not comprehensive of all events that happened in Ivory Coast in the monitored 5-month period, and this explains the low precision of both methods. 
This is the reason why we did not try to find the optimal values of the parameters to maximize Precision and Recall, but instead set such values based on the general criteria to find unusual-crowds only in around 10-15\% of the hours.
However, the list in Table \ref{tb:exp-events} gives a good basis for comparison and shows that our method is 10 times more precise than GAT.

We perform another comparison with MOV, where the authors introduce the concept of moving clusters.
Extracting moving clusters is equivalent to run our method with the parameter $\epsilon_p$ (the probability of a user to be committed) and $\epsilon_{si}$ (the similarity threshold between the mobility profile and the trajectories in the event) to 1. With these parameter settings the algorithm was not able to find any moving clusters.
This is due to the fact that our method is able to handle the spatial and temporal sparsity of CDR data, while the MOV method is designed to work with GPS trajectories.

\vpara{Time-series.} 
We report the time series of the numbers of unusual events and gatherings detected with different input parameter settings by using our method and the GAT algorithm in Figures \ref{figsub:exp-weeks0-ue} and \ref{figsub:exp-weeks0-gat}. 
We try to match the same parameters we used in our method. For the minimum lifetime as well as the minimum number of objects that should belong to a cluster, we choose the same values adopted in the study of the effectiveness of our method ($\epsilon_{lt}$ = 4 and $\epsilon_{n}$ = 20). The rest of the input parameters of the algorithm to detect gatherings are the following: $d$ is the minimum distance necessary to connect clusters detected in two consecutive time snapshots; $k_p$ is the minimum number of time snapshots required to consider an user as a participant; $m_p$ is the minimum number of participants to create a gathering. For these input parameters, we tried different enumerations to span the full admissible ranges. 
As it is possible to see, the number of detected gatherings is very high even if the parameters are chosen to be very restrictive. All the graphs show a daily trend, demonstrating that this method is not able to find unusual events as we propose in this paper. Indeed, we can detect a large number of gatherings every day, that might not correspond to specific unusual events.

To further evaluate our discoveries, we check the total communication volumes and the specific antenna activities.  
We can clearly see that between Dec. 06 and Dec. 18, in Figure \ref{figsub:exp-weeks0-times} there exist periodic patterns on each day without obvious peak values corresponding to the discovery of crowds---\textbf{anniversary of Felix death} on Dec. 07 and \textbf{Parliament election} on Dec. 11. 
Furthermore, the events on the day of parliament election involved five antennas. 
Their communication activities are plotted in Figure \ref{figsub:exp-weeks0-bids}. 
Obviously there do not exist correlations between corresponding antenna activities and our unusual crowd/event output.
These two regular and stable time series of communication activities further confirm the effectiveness of our problem design. These examples show that detecting unusual events is a complex task, which cannot be easily accomplished by looking at outliers in call time series. Thus, methods like \cite{DKE13:Zheng,ICDM12:Zheng} are not directly applicable.

\begin{figure}[!t]
\centering
\subfigure[\scriptsize Detected unusual events]{
\label{figsub:ourSpatial}
\includegraphics[width=2.8in]{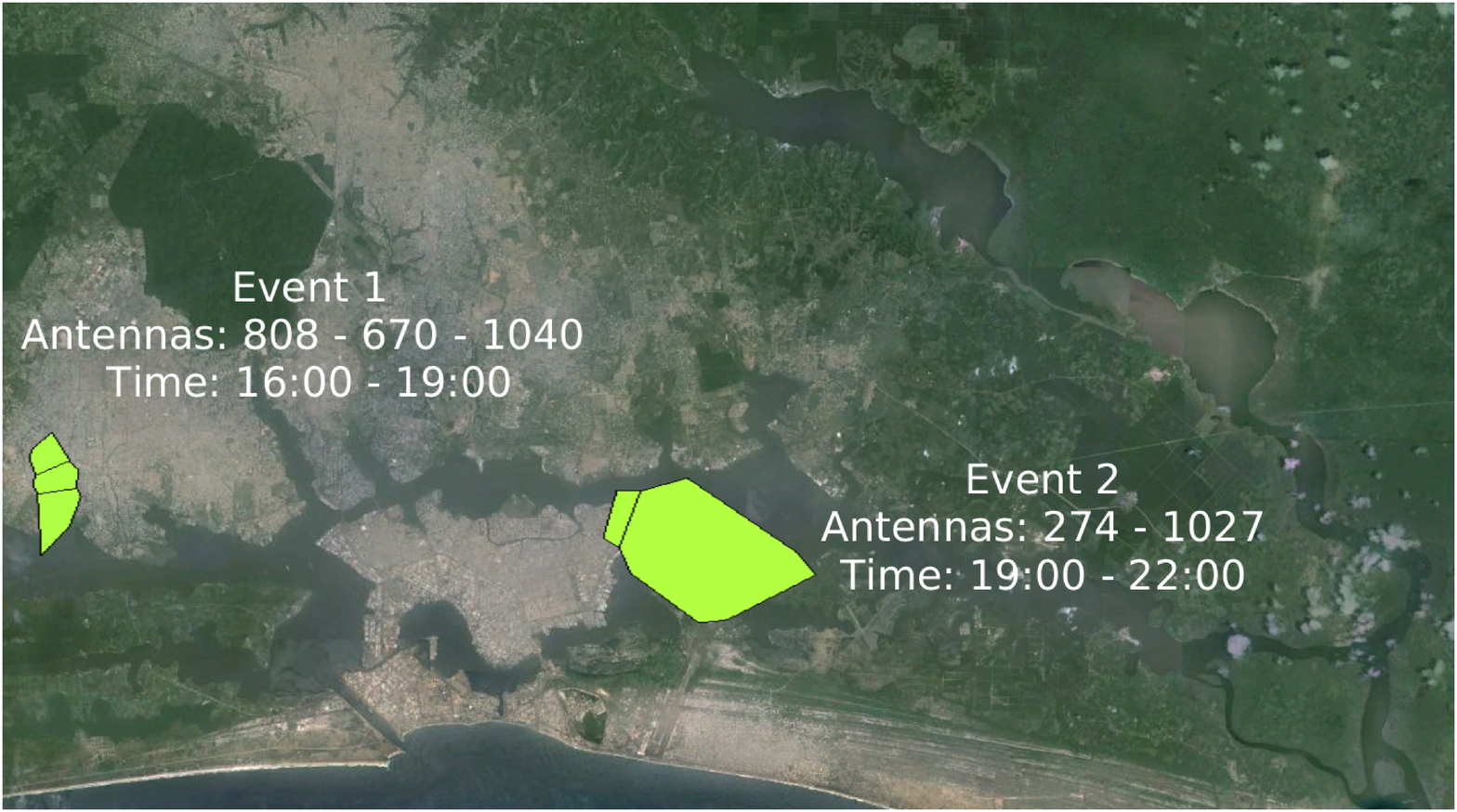}
}
\subfigure[\scriptsize Detected gatherings]{
\label{figsub:icdeSpatial}
\includegraphics[width=2.8in]{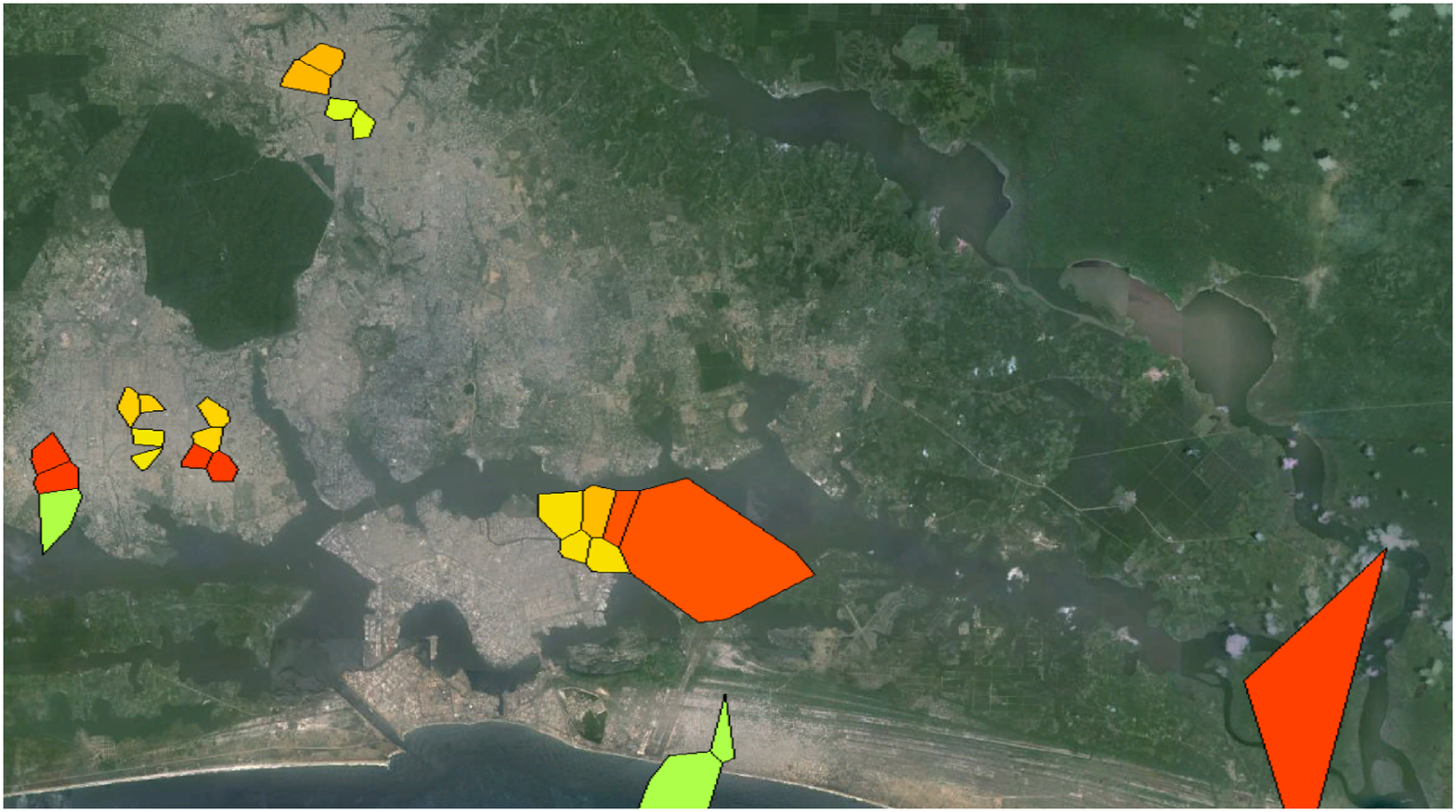}
}
\caption{\label{fig:spatialComparison} The unusual events (a) by our methods and the gatherings (b) by GAT detected on December 11th. Colors range from green to red as function of the number of detected participants}
\end{figure}

\vpara{Spatial distributions.}
Another comparison performed against GAT regards the spatial distribution of the detected events/gatherings.
For both methods, we select the results obtained on December 11th. For the GAT method we select the results with lowest number of detected gatherings. 
In Figures \ref{figsub:ourSpatial} and \ref{figsub:icdeSpatial} we report the detected events and detected gatherings respectively. Notice that a Voronoi tessellation has been applied in order to associate a covering area to each antenna.
Our method detect 2 events, \emph{Event 1} (left) covers 3 antennas and it lasts for 4 hours. \emph{Event 2} (right) covers 2 antennas and also this one lasts for 4 hours.
For the same day, the algorithm of Zheng et al. detects many gatherings (25) occurring in different places of the city.
This is probably due to the fact that the typical mobility profiles of the users are not taken into account in the process and recurring events and unusual are both detected.
Moreover, if we consider the lifetime of gatherings occurring in the same locations of our events, we notice that it is generally longer. For example, a gathering, covering the same antennas of \emph{Event 1}, lasts for 14 hours.
Another characteristic of the gatherings is that they happen in the same location at different times. Instead, in our model, we define a method to consider those as one large event.
In summary, with our method it is possible to identify events that occur occasionally in a precise zone of the city and happen in a precise period of time, while the other method detects several events without any distinctions between the periodical ones and the unusual ones.

\subsection{Efficiency and Parameters}
\label{sec:efficeity-para}

Our algorithms are implemented in Python 2.7.5, and all experiments were performed on a laptop running Windows 7 with Intel(R) Core(TM) i7-2720QM CPU@2.20GHz (2 cores) and 8GB memory. 
All related experiments are running on the first two-week dataset, which contains about two million CDR historic data.
We simulate each experiment with specific parameter setting for 100 times to get both the average running time and standard deviation.
In general, the algorithms for our specific problems are efficient, in the fact that it only takes about 30 seconds to two minutes on two million CDR data.
Furthermore, the execution of our methods is stable among the different runs.

We evaluate the influence of parameter setting on the number of detected unusual crowds and discuss the guidelines for determining parameter settings. 
We find that the algorithm is particularly sensitive to $\epsilon_{lt}$, $\epsilon_{p}$, and $\epsilon_{si}$. 
$\epsilon_{lt}$ is indicative of the duration of moving crowds. 
Based on the definition of \textit{commitment}, a larger $\epsilon_{p}$ can produce more compact crowds, which have a much higher probability to be unusual events. 
Finally, the lower \textit{similarity} $\epsilon_{si}$ threshold between regular mobility profiles and specific trajectory we set, the more crowds will consist of people whose mobility behavior differs from their typical profiles. 

We would like to point out that there is not an unique way to optimally select the values of the parameters, as this strongly depends on the end-user application. For instance, if a city manager is interested in monitoring visitors to a museums, she might set different values of the $\epsilon_{lt}$ and $\epsilon_{p}$, compared to the monitoring on a protest. 
We develop a visual analytics tool described in Section \ref{sec:tool} to help end users explore and test the detected results under different parameter settings for different applications.


%% file: tool.tex
\section{Visual Analytics Prototype System}
\label{sec:tool}

\begin{figure}[!t]
\centering
\includegraphics[width=1\linewidth]{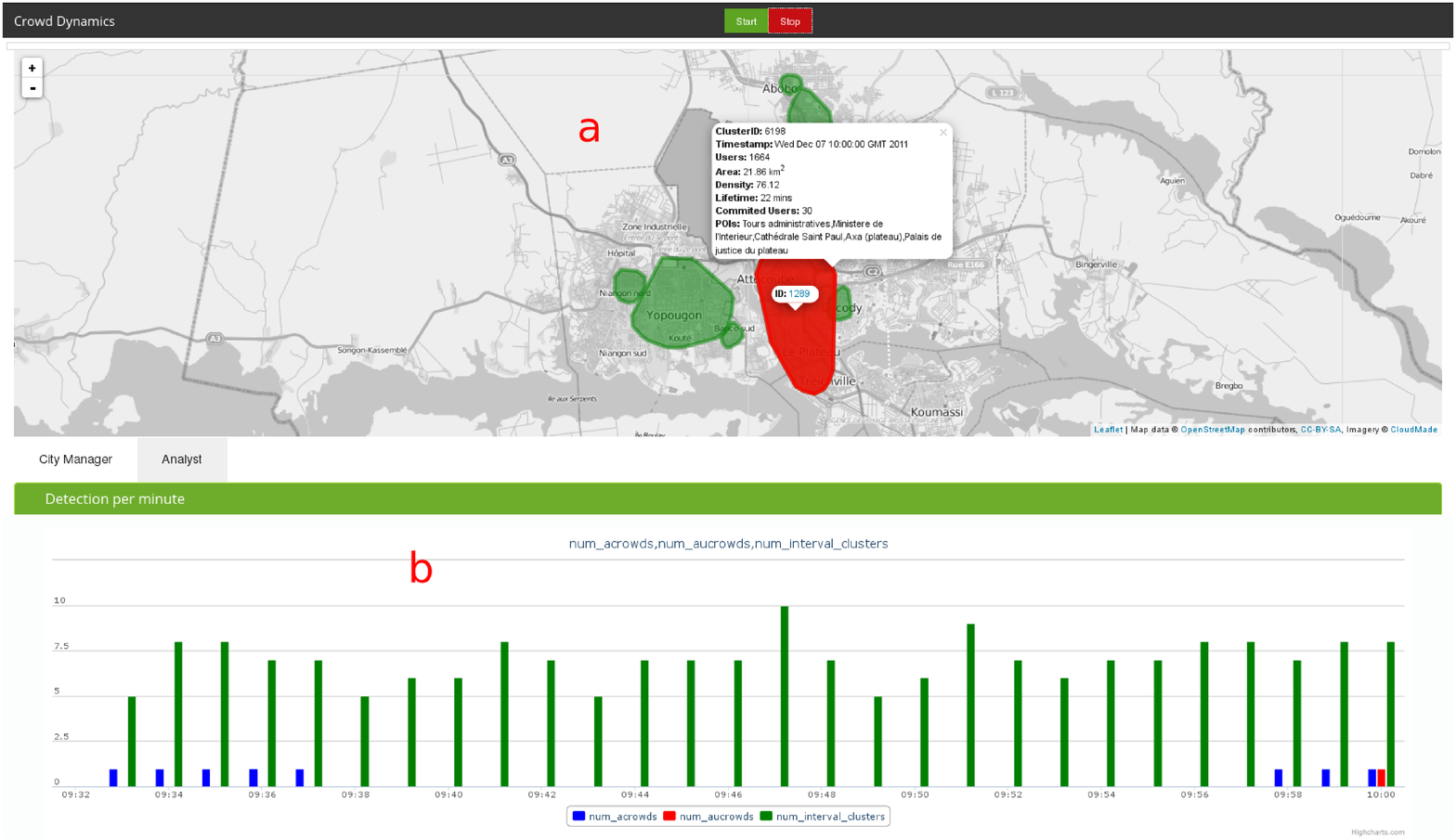}
\caption{\label{fig:Realtime-Overview}Real-time view of city map and statistics.}
\end{figure}

We have developed a visual analytics system to support the exploration of the unusual crowd events based on the proposed framework. 
The system allows end users---such as analysts and city managers---to analyze the formation and evolution of crowds, and study the impact of different parameters on the obtained results, heuristically suggesting possible changes to get more meaningful results depending on the desired application. 
The interface consists mainly of two components: the map overview of the observed city (Figure \ref{fig:Realtime-Overview} (a)) and the statistics of users, crowds, and events (Figure \ref{fig:Realtime-Overview} (b)). 

\vpara{Map view.} 
In the map, the system visualizes the latest clusters, crowds and unusual events detected in the form of polygons as shown in Figure \ref{fig:Realtime-Overview} (a). 
The polygons are the convex hulls of the location updates of the users belonging to one of the aforementioned groups. 
The clusters detected by the system at a given timestamp are visualized on the map as green polygons. 
On mouse-over, the UI shows a pop-up window with the cluster attributes, including 1) timestamp---the timestamp when the cluster was detected, 2) \#users---the number of users that are a part of the cluster, 3) area---the area covered by the Cluster polygon in square kilometers, 4) density---the ratio between the number of users in the Cluster
and its area, and 5) POIs---the list of Points Of Interest located within the cluster area.  
The detected Crowds and the Unusual events are visualized on the map as blue and red polygons, respectively. 
Similar to the cluster visualization, a set of properties can be shown in a pop-up window. 

\begin{figure}[!t]
\centering
\includegraphics[width=1\linewidth]{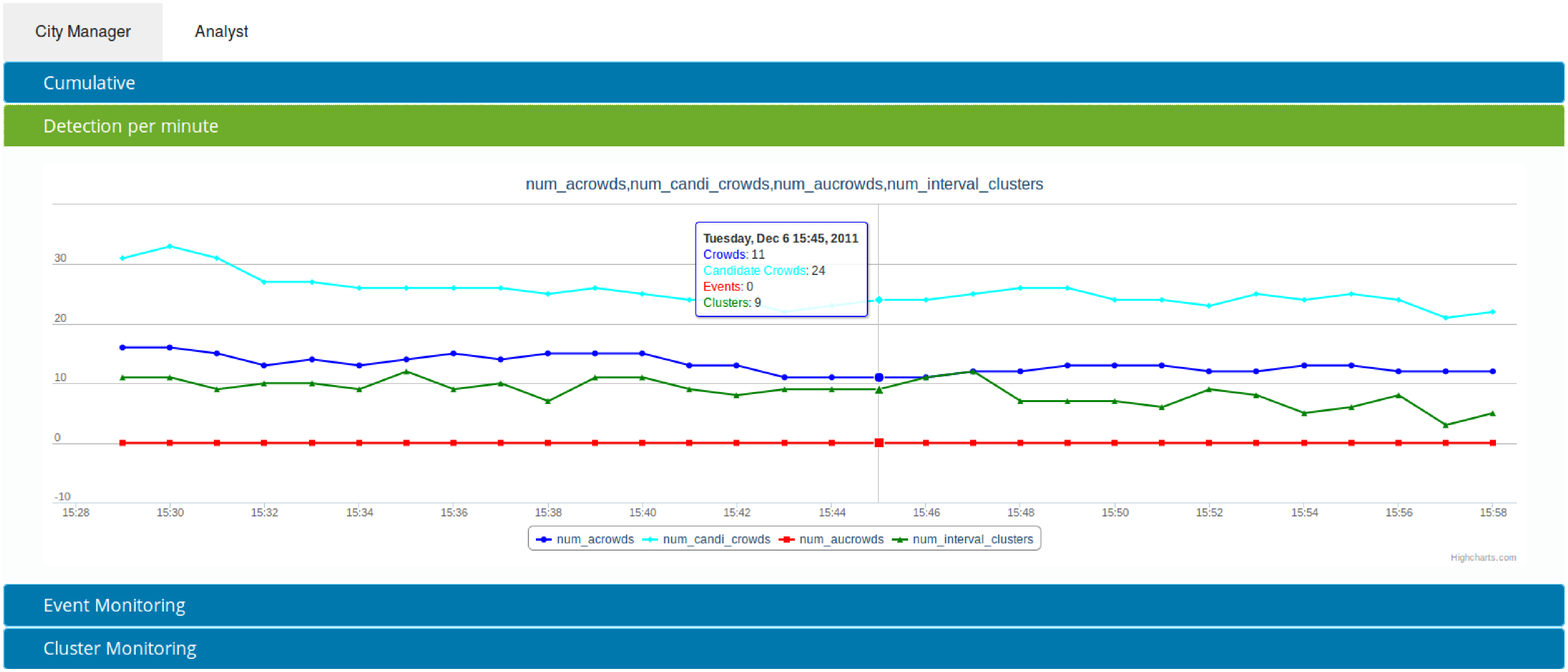}
\caption{\label{fig:Analyst-Statistics}Analyst statistics view}
\end{figure}

\vpara{\label{sub:Statistics-View}Statistics view.}
We envision the tool to be used by two different types of actors: the Analyst, which is in charge of setting up the analytics system, and the city manager, who has to take actions based on the identified unusual crowd events. 
The City manager tab contains a single time-series graph representing the number of Clusters, Crowds, Unusual Events detected at every timestamp as displayed in Figure \ref{fig:Realtime-Overview} (b). 
This tab contains the most crucial outcome of the analytics performed and it provides an intuitive way to represent the most recent mobility patterns of the city.
The Analyst tab contains a richer set of statistics in Figure \ref{fig:Analyst-Statistics}. 
In order to make efficient use of available space we fit the graphs into collapsible panels into groups of semantically relevant statistics, including 1) Cumulative---the cumulative trends of the detected clusters, crowds and unusual events, 2) Detection per minute---the time-series of the number of clusters, candidate crowds, crowds, and unusual events, 3) Event monitoring---the time-series of maximum and minimum value of lifetime, number of committed users, total number of users, and similarity of the candidate crowds, 4) Cluster monitoring---the time-series of the maximum size of the detected clusters, and their minimum spatial radius. 
In addition, a red dashed-line corresponding to each parameter is shown to depict parameter efficacy (e.g. event monitoring, cluster monitoring). 
This allows the Analyst to understand the role of each parameter on the obtained results and to set the most appropriate parameters for the specific application in scope.

%% file: related.tex
\section{Related Work}
\label{sec:related}
The availability of mobility data has offered researchers the opportunity to analyze both individual's and group's moving behaviors. 
In \cite{denseareas,denseareas2}, the authors defined a methodology to extract dense areas in spatio-temporal databases, thus identifying where and when dense areas of mobile objects appear.
A similar definition was proposed in \cite{movingClusters}, where the authors introduced the concept of \emph{moving clusters}. 
Following these ideas, several group and cluster mobility pattern models have been proposed~\cite{Donato:DMKD14,Mobile:TKDE11,TKDD10:anomaly,Convoy:VLDB08}. 
For instance, in \cite{flock1,flock2,flock3,flockrebecca} the concept of \emph{flock} is widely investigated.
A flock is a group of objects that travel together within a disc of some user-specified size for at least $k$ consecutive timestamps. 
The main limit of this model is that a simple circular shape does not reflect natural grouping in reality.
A marginal different concept \emph{convoy} was introduced in \cite{Convoy:VLDB08,convoy2}, where a density-based clustering is adopted instead of the radius of the disk. 
Li at al. \cite{swarm} proposed a more general type of trajectory patterns: \emph{swarm}. 
The swarm is a cluster of moving objects that lasts for at least $k$ timestamps, possibly non-consecutive.
Another group mobility model was introduced in \cite{Gathering:ICDE13}, called \emph{gathering}, whose 
novelty regards the introduction of the concept of commitment. 

Other interesting works dealing with the detection of anomalies in city traffic flow are presented in \cite{Aditya:DMKD14,DKE13:Zheng,ICDM12:Zheng}. In \cite{DKE13:Zheng}, the authors use  likelihood ratio test statistic (LRT) on GPS trajectories of taxis to detect traffic flow anomalies.
Transportation model detection problem is studied in mobile phone data and GIS data \cite{MobileGIS:GIS'11}. 
A passive route sensing framework is introduced to monitor users' significant driving routes with low-power sensors in mobile phones~\cite{Cecilia:SenSys2014}. 
However, these works do not address the problem detecting unusual events considering people mobility but are more focused on traffic flow analysis through an aggregation of the information. On the contrary, in this paper we are interested in detecting such events that involve a large number of people whose current mobility differs from their typical one.

All the above works are designed and tested on high-resolution trajectory data, such as the one provided by GPS systems.
Low-resolution location data collected from telecommunication operators, on the other hand, is much more pervasive resulting in a much larger sample of the population being monitored, see \cite{Calabrese:14,Isaacman:2012:HMM:2307636.2307659,Calabrese_Pervasive2010,Cecilia:ICWSM2014}.
In this paper, we propose a new method to mine coarse grain mobile phone data (in the form of CDR) to detect unusual crowd events.
Indeed, the aim of our work is to detect events that involve a large number of people performing unusual activities. 
To do so, we compute a similarity between the mobility profile of the users and their trajectories in group pattern. This extends, thus, the concept of commitment since users need to be committed and have trajectories that differ from the ones in their mobility profiles. 
Our method however is able to identify moving events that span several locations over time, and involve a subset of committed users, something that could not be detected by using the methods in \cite{Francesco:SOCIALCOM11,PervasiveComp13}.

%% file: conclusion.tex
\section{Conclusion and Future Work}
\label{sec:conclusion}
In this paper, we formally define the problem of inferring unusual crowd events from mobility data. 
Previous work on event detection is limited on inferring the usual event from the fine-grained GPS data.
Our problem definition differs by characterizing the unusual crowd events and presenting a new methodology to extract them from coarse-grained CDR data.
The main contributions of this paper w.r.t. existing methods are the ability to analyze temporally and spatially sparse data as CDRs and the definition of a subclass of events which are unusual to its attendees. 
Our experimental results demonstrate the effectiveness of our method in a real-world mobile data. 


Despite the promising results of the present work, there is still much room left for future work. 
First, while this proposed method relies on Visual Analytics to help end users set parameters, we are planning to design algorithms to determine   parameters for specific applications of interest as well as an optimization procedure for evaluation metrics.
Moreover, we are working toward combining mobile and social media data together to detect unusual events. 
In doing so, we have the potential to detect and monitor crowding activities in real time, and eventually yield a better and smarter planet.

%% file: crowd-main-arxiv.bbl
\begin{thebibliography}{10}

\bibitem{Donato:DMKD14}
A.~Appice and D.~Malerba.
\newblock Leveraging the power of local spatial autocorrelation in geophysical
  interpolative clustering.
\newblock {\em DMKD}, 28(5-6):1266--1313, Sept. 2014.

\bibitem{flock1}
M.~Benkert, J.~Gudmundsson, F.~Hübner, and T.~Wolle.
\newblock Reporting flock patterns.
\newblock {\em Computational Geometry}, 41(3):111 -- 125, 2008.

\bibitem{SEA:Netmob2015}
F.~Calabrese, G.~Di~Lorenzo, G.~McArdle, F.~Pinelli, and E.~Van~Lierde.
\newblock Real-time social event analytics.
\newblock In {\em Netmob '15}, 2015.

\bibitem{Calabrese:14}
F.~Calabrese, L.~Ferrari, and V.~Blondel.
\newblock Urban sensing using mobile phones network data: A survey of research.
\newblock {\em ACM Comput. Surv.}, 2014.

\bibitem{Calabrese_Pervasive2010}
F.~Calabrese, F.~Pereira, G.~Lorenzo, L.~Liu, and C.~Ratti.
\newblock The geography of taste: Analyzing cell-phone mobility and social
  events.
\newblock In {\em Pervasive Computing'10}, pages 22--37. 2010.

\bibitem{ICDM12:Zheng}
S.~Chawla, Y.~Zheng, and J.~Hu.
\newblock Inferring the root cause in road traffic anomalies.
\newblock In {\em IEEE ICDM'12}, pages 141--150, 2012.

\bibitem{Dong:KDD14}
Y.~Dong, Y.~Yang, J.~Tang, Y.~Yang, and N.~V. Chawla.
\newblock Inferring user demographics and social strategies in mobile social
  networks.
\newblock In {\em KDD '14}, pages 15--24. ACM, 2014.

\bibitem{DBSCAN:KDD96}
M.~Ester, H.~peter Kriegel, J.~S, and X.~Xu.
\newblock A density-based algorithm for discovering clusters in large spatial
  databases with noise.
\newblock In {\em ACM SIGKDD'96}, pages 226--231, 1996.

\bibitem{Cecilia:ICWSM2014}
P.~Georgiev, A.~Noulas, and C.~Mascolo.
\newblock The call of the crowd: Event participation in location-based social
  services.
\newblock In {\em ICWSM'14}, 2014.

\bibitem{Trajectory:KDD07}
F.~Giannotti, M.~Nanni, F.~Pinelli, and D.~Pedreschi.
\newblock Trajectory pattern mining.
\newblock In {\em ACM SIGKDD'07}, pages 330--339, New York, NY, USA, 2007. ACM.

\bibitem{flock3}
J.~Gudmundsson and M.~van Kreveld.
\newblock Computing longest duration flocks in trajectory data.
\newblock In {\em ACM GIS'06}, pages 35--42, New York, NY, USA, 2006. ACM.

\bibitem{denseareas}
M.~Hadjieleftheriou, G.~Kollios, D.~Gunopulos, and V.~J. Tsotras.
\newblock On-line discovery of dense areas in spatio-temporal databases.
\newblock In {\em Proc. of the 7th International Conference on Advances in
  Spatial and Temporal Databases}, SSTD'03, pages 306--324, 2003.

\bibitem{Isaacman:2012:HMM:2307636.2307659}
S.~Isaacman, R.~Becker, R.~C\'{a}ceres, M.~Martonosi, J.~Rowland,
  A.~Varshavsky, and W.~Willinger.
\newblock Human mobility modeling at metropolitan scales.
\newblock In {\em Proceedings of the 10th International Conference on Mobile
  Systems, Applications, and Services}, MobiSys '12, pages 239--252, New York,
  NY, USA, 2012. ACM.

\bibitem{denseareas2}
C.~S. Jensen, D.~Lin, B.~C. Ooi, and R.~Zhang.
\newblock Effective density queries on continuously moving objects.
\newblock In {\em IEEE ICDE '06}, pages 71--, Washington, DC, USA, 2006.

\bibitem{convoy2}
H.~Jeung, H.~T. Shen, and X.~Zhou.
\newblock Convoy queries in spatio-temporal databases.
\newblock In {\em IEEE ICDE '08}, pages 1457--1459, Washington, DC, USA, 2008.
  IEEE Computer Society.

\bibitem{Convoy:VLDB08}
H.~Jeung, M.~L. Yiu, X.~Zhou, C.~S. Jensen, and H.~T. Shen.
\newblock Discovery of convoys in trajectory databases.
\newblock {\em Proc. VLDB Endow.}, 1(1):1068--1080, Aug. 2008.

\bibitem{movingClusters}
P.~Kalnis, N.~Mamoulis, and S.~Bakiras.
\newblock On discovering moving clusters in spatio-temporal data.
\newblock In {\em Proc. of the 9th International Conference on Advances in
  Spatial and Temporal Databases}, SSTD'05, pages 364--381, Berlin, Heidelberg,
  2005. Springer-Verlag.

\bibitem{swarm}
Z.~Li, B.~Ding, J.~Han, and R.~Kays.
\newblock Swarm: mining relaxed temporal moving object clusters.
\newblock {\em Proc. VLDB Endow.}, 3(1-2):723--734, 2010.

\bibitem{Mobile:TKDE11}
E.~H.-C. Lu, V.~S. Tseng, and P.~S. Yu.
\newblock Mining cluster-based temporal mobile sequential patterns in
  location-based service environments.
\newblock {\em IEEE Trans. on Knowl. and Data Eng.}, 23(6):914--927, June 2011.

\bibitem{Cecilia:SenSys2014}
S.~Nawaz and C.~Mascolo.
\newblock Mining users' significant driving routes with low-power sensors.
\newblock In {\em ACM SenSys '14}, pages 236--250. ACM, 2014.

\bibitem{DKE13:Zheng}
L.~X. Pang, S.~Chawla, W.~Liu, and Y.~Zheng.
\newblock On detection of emerging anomalous traffic patterns using gps data.
\newblock {\em Data Knowl. Eng.}, 87:357--373, 2013.

\bibitem{netmobevents}
P.~Paraskevopoulos, T.-C. Dinh, Z.~Dashdorj, T.~Palpanas, and L.~Serafini.
\newblock Identification and characterization of human behavior patterns from
  mobile phone data.
\newblock In {\em International Conference on the Analysis of Mobile Phone
  Datasets (NetMob'13)}, 2013.

\bibitem{MobileGIS:GIS'11}
L.~Stenneth, O.~Wolfson, P.~S. Yu, and B.~Xu.
\newblock Transportation mode detection using mobile phones and gis
  information.
\newblock In {\em GIS '11}, pages 54--63, New York, NY, USA, 2011. ACM.

\bibitem{KDD13:conformity}
J.~Tang, S.~Wu, and J.~Sun.
\newblock Confluence: conformity influence in large social networks.
\newblock In {\em ACM SIGKDD '13}, pages 347--355, New York, NY, USA, 2013.
  ACM.

\bibitem{Aditya:DMKD14}
A.~Telang, P.~Deepak, S.~Joshi, P.~Deshpande, and R.~Rajendran.
\newblock Detecting localized homogeneous anomalies over spatio-temporal data.
\newblock {\em DMKD}, 28(5-6):1480--1502, 2014.

\bibitem{Francesco:SOCIALCOM11}
V.~A. Traag, A.~Browet, F.~Calabrese, and F.~Morlot.
\newblock Social event detection in massive mobile phone data using
  probabilistic location inference.
\newblock In {\em SocialCom'11}, pages 625--628. IEEE, 2011.

\bibitem{Car:KDD11}
R.~Trasarti, F.~Pinelli, M.~Nanni, and F.~Giannotti.
\newblock Mining mobility user profiles for car pooling.
\newblock In {\em ACM SIGKDD'11}, pages 1190--1198, New York, NY, USA, 2011.
  ACM.

\bibitem{flock2}
M.~R. Vieira, P.~Bakalov, and V.~J. Tsotras.
\newblock On-line discovery of flock patterns in spatio-temporal data.
\newblock In {\em ACM GIS'09}, pages 286--295, New York, NY, USA, 2009. ACM.

\bibitem{flockrebecca}
M.~Wachowicz, R.~Ong, C.~Renso, and M.~Nanni.
\newblock Finding moving flock patterns among pedestrians through collective
  coherence.
\newblock {\em International Journal of Geographical Information Science},
  25(11):1849--1864, 2011.

\bibitem{PervasiveComp13}
A.~Witayangkurn, T.~Horanont, Y.~Sekimoto, and R.~Shibasaki.
\newblock Anomalous event detection on large-scale gps data from mobile phones
  using hidden markov model and cloud platform.
\newblock In {\em ACM UbiComp '13 Adjunct}, pages 1219--1228, New York, NY,
  USA, 2013. ACM.

\bibitem{TKDD10:anomaly}
M.~Wu, C.~Jermaine, S.~Ranka, X.~Song, and J.~Gums.
\newblock A model-agnostic framework for fast spatial anomaly detection.
\newblock {\em ACM Trans. Knowl. Discov. Data}, 4(4):20:1--20:30, Oct. 2010.

\bibitem{Gathering:ICDE13}
K.~Zheng, Y.~Zheng, N.~J. Yuan, and S.~Shang.
\newblock On discovery of gathering patterns from trajectories.
\newblock In {\em IEEE ICDE'13}, pages 242--253, Washington, DC, USA, 2013.

\end{thebibliography}
